\documentclass[twocolumn,trackinematic centrehanges]{aastex701}

\newcommand\kms{\ensuremath{{\rm km}~{\rm s}^{-1}}\xspace}
\newcommand\um{\ensuremath{{\rm \mu m}}\xspace}
\newcommand{\ha}{\ensuremath{\mathrm{H}\alpha}\xspace}
\newcommand{\paa}{\ensuremath{\mathrm{Pa}\alpha}\xspace}
\newcommand{\h}[2]{\ensuremath{\mathrm{H_2\ 1-}\mathrm{0\ #1(#2)}}\xspace}
\newcommand{\hs}[1]{\h{S}{#1}}

\newcommand{\ho}[1]{\h{O}{#1}}
\newcommand{\msun}{\ensuremath{\mathrm{M}_\odot}\xspace}
\newcommand{\msunpyr}{\ensuremath{\mathrm{M}_\odot~\mathrm{yr}^{-1}}\xspace}
\usepackage{xspace}
\usepackage{color,soul}
\usepackage{amsmath}
\usepackage{CJK}
\usepackage{siunitx}
\usepackage{hyperref}

\begin{document}
\title{Mapping Gas Accretion and Stellar Kinematics to Sub-kiloparsec Scales in NGC~4696 with JWST/NIRSpec}
\shorttitle{Kinematics of the Gas and Stars in NGC~4696 with JWST/NIRSpec}
\shortauthors{Marquis et al.}

\begin{CJK}{UTF8}{}
\CJKfamily{mj}

\correspondingauthor{Mathieu Marquis}
\email{mathieu.marquis@umontreal.ca}

\author[0009-0008-7156-4678]{Mathieu Marquis}
\affiliation{D\'{e}partement de Physique, Universit\'{e} de Montr\'{e}al, Succ. Centre-Ville,
Montr\'{e}al, Qu\'{e}bec, H3C 3J7, Canada}
\email{...}

\author[0000-0001-7271-7340]{Julie Hlavacek-Larrondo}
\affiliation{D\'{e}partement de Physique, Universit\'{e} de Montr\'{e}al, Succ. Centre-Ville, Montr\'{e}al, Qu\'{e}bec, H3C 3J7, Canada}
\email{...}

\author[0009-0000-4104-9909]{Olivia Pereira}
\affiliation{D\'{e}partement de Physique, Universit\'{e} de Montr\'{e}al, Succ. Centre-Ville,
Montr\'{e}al, Qu\'{e}bec, H3C 3J7, Canada}
\email{...}

\author[0000-0003-4701-8497]{Michael Reefe}
\affiliation{Kavli Institute for Astrophysics and Space Research,
Massachusetts Institute of Technology,
77 Massachusetts Avenue, Cambridge, MA 02139, USA}
\email{}

\author[0000-0002-3173-1098]{Hyunseop Choi (최현섭)}
\affiliation{Department of Astronomy, University of Michigan, 1085 S. University, Ann Arbor, MI 48109, USA}
\email{...}

\author[0000-0003-2405-7258]{Jorge Barrera-Ballesteros}
\affiliation{Instituto de Astronom\'{i}a, Universidad Nacional Aut\'{o}noma de M\'{e}xico,
A.P. 70-264, 04510 CDMX, M\'{e}xico}
\email{}

\author[0000-0002-2478-5119]{Benjamin Vigneron}
\affiliation{D\'{e}partement de Physique, Universit\'{e} de Montr\'{e}al,
Succ. Centre-Ville, Montr\'{e}al, Qu\'{e}bec, H3C 3J7, Canada}
\email{}

\author[0000-0001-5880-0703]{Ming Sun}
\affiliation{Department of Physics and Astronomy, University of Alabama in Huntsville,
301 Sparkman Drive, Huntsville, AL 35899, USA}
\email{}

\author[0000-0003-1398-5542]{Rebecca E. A. Canning}
\affiliation{Institute of Cosmology and Gravitation, University of Portsmouth,
Dennis Sciama Building, Portsmouth, PO1 3FX, UK}
\email{}

\author[0000-0001-6495-7731]{Gregory Taylor}
\affiliation{Department of Physics and Astronomy, University of New Mexico,
Albuquerque, NM 87131, USA}
\email{}

\author[0000-0003-0475-9375]{Lo\"ic Albert}
\affiliation{Institut Trottier de recherche sur les exoplan\`etes and D\'epartement de Physique, Universit\'e de Montr\'eal, 1375 Avenue Th\'er\`ese-Lavoie-Roux, Montr\'eal, QC H2V 0B3, Canada}
\email{placeholder}

\author[0000-0003-2388-8172]{Francesco D'Eugenio}
\affiliation{Kavli Institute for Cosmology, University of Cambridge,
Madingley Road, Cambridge, CB3 0HA, UK}
\affiliation{Cavendish Laboratory, University of Cambridge,
19 JJ Thomson Avenue, Cambridge, CB3 0HE, UK}
\email{}

\author[0000-0002-2808-0853]{Megan Donahue}
\affiliation{Department of Physics and Astronomy, Michigan State University,
East Lansing, MI 48824, USA}
\email{}

\author[0000-0002-9378-4072]{Andrew C. Fabian}
\affiliation{Institute of Astronomy, University of Cambridge,
Madingley Road, Cambridge, CB3 0HA, UK}
\email{}

\author[0000-0003-4503-6333]{Gary J. Ferland}
\affiliation{Department of Physics and Astronomy, University of Kentucky,
505 Rose Street, Lexington, KY 40506, USA}
\email{}

\author[0000-0001-8608-0408]{John S. Gallagher}
\affiliation{Department of Physics and Astronomy, Macalester College,
1600 Grand Avenue, Saint Paul, MN 55105, USA}
\email{}

\author[0000-0002-7326-5793]{Marie-Lou Gendron-Marsolais}
\affiliation{D\'{e}partement de Physique, de G\'{e}nie Physique et d'Optique,
Universit\'{e} Laval, Qu\'{e}bec, QC G1V 0A6, Canada}
\email{}

\author[0000-0002-2421-1350]{Pierre Guillard}
\affiliation{Sorbonne Universit\'{e}, CNRS, UMR 7095, Institut d'Astrophysique de Paris,
98bis bd Arago, 75014 Paris, France}
\email{}

\author[0000-0002-3680-5420]{Minghao Guo (郭明浩)}
\affiliation{Department of Astrophysical Sciences, Princeton University, Princeton, NJ 08540, USA}
\email{...}

\author[0000-0001-5600-0534]{Nina Hatch}
\affiliation{School of Physics and Astronomy, University of Nottingham,
University Park, Nottingham, NG7 2RD, UK}
\email{}

\author[0000-0002-4460-9892]{Ralf Kotulla}
\affiliation{Department of Astronomy, University of Wisconsin--Madison,
475 N. Charter Street, Madison, WI 53706, USA}
\email{}

\author[0000-0001-5262-6150]{Yuan Li}
\affiliation{Department of Astronomy, University of Massachusetts,
Amherst, MA 01003, USA}
\email{}

\author[0000-0002-4985-3819]{Roberto Maiolino}
\affiliation{Kavli Institute for Cosmology, University of Cambridge,
Madingley Road, Cambridge, CB3 0HA, UK}
\affiliation{Cavendish Laboratory, University of Cambridge,
19 JJ Thomson Avenue, Cambridge, CB3 0HE, UK}
\affiliation{Department of Physics and Astronomy, University College London,
Gower Street, London WC1E 6BT, UK}
\email{}

\author[0000-0003-2475-124X]{Allison Man}
\affiliation{Department of Physics \& Astronomy, University of British Columbia,
6224 Agricultural Road, Vancouver, BC V6T 1Z1, Canada}
\email{}

\author[0000-0001-5226-8349]{Michael A. McDonald}
\affiliation{Kavli Institute for Astrophysics and Space Research,
Massachusetts Institute of Technology,
77 Massachusetts Avenue, Cambridge, MA 02139, USA}
\email{}

\author[0000-0002-2622-2627]{Brian R. McNamara}
\affiliation{Department of Physics and Astronomy, University of Waterloo,
200 University Avenue West, Waterloo, ON N2L 3G1, Canada}
\email{}

\author[0000-0001-6638-4324]{Valeria Olivares}
\affiliation{Departamento de F\'{i}sica, Universidad de Santiago de Chile,
Av. Victor Jara 3659, Santiago 9170124, Chile}
\affiliation{Center for Interdisciplinary Research in Astrophysics and Space Exploration (CIRAS),
Universidad de Santiago de Chile, Santiago 9170124, Chile}
\email{}

\author[0009-0003-0932-2487]{Marine Prunier}
\affiliation{D\'{e}partement de Physique, Universit\'{e} de Montr\'{e}al,
Succ. Centre-Ville, Montr\'{e}al, Qu\'{e}bec, H3C 3J7, Canada}
\affiliation{Max-Planck-Institut f{\"u}r Astronomie, K{\"o}nigstuhl 17, D-69117 Heidelberg, Germany}
\email{}

\author[0000-0002-1510-4860]{Christopher S. Reynolds}
\affiliation{Department of Astronomy, University of Maryland,
College Park, MD 20742-2421, USA}
\affiliation{Joint Space Science Institute (JSI), University of Maryland,
College Park, MD 20742-2421, USA}
\email{}

\author[0000-0003-2001-1076]{Carter Rhea}
\affiliation{Dragonfly Focused Research Organization, 150 Washington Avenue, Santa Fe, 87501, NM, USA}
\affiliation{Centre de Recherche en Astrophysique du Qu\'{e}bec (CRAQ), Qu\'{e}bec, QC G1V 0A6, Canada}
\email{}

\author[0000-0001-7597-270X]{Annabelle Richard-Laferri\`{e}re}
\affiliation{Institute of Astronomy, University of Cambridge,
Madingley Road, Cambridge, CB3 0HA, UK}
\email{}

\author[0000-0001-5208-649X]{Helen R. Russell}
\affiliation{School of Physics and Astronomy, University of Nottingham,
University Park, Nottingham, NG7 2RD, UK}
\email{}

\author[0000-0001-9633-5750]{Philippe Salom\'{e}}
\affiliation{LERMA, Observatoire de Paris, PSL Research University, CNRS,
Sorbonne Universit\'{e}, 75014 Paris, France}
\email{}

\author[0000-0001-8176-7665]{Prathamesh Tamhane}
\affiliation{Department of Physics and Astronomy, University of Alabama in Huntsville,
301 Sparkman Drive, Huntsville, AL 35899, USA}
\email{}

\author[0000-0001-5223-1888]{Auriane Thilloy}
\affiliation{D\'{e}partement de Physique, Universit\'{e} de Montr\'{e}al,
Succ. Centre-Ville, Montr\'{e}al, Qu\'{e}bec, H3C 3J7, Canada}
\email{}

\author[0000-0002-5445-5401]{Grant R.~Tremblay}
\affiliation{Harvard-Smithsonian Center for Astrophysics,
60 Garden Street, Cambridge, MA, USA}
\email{}

\author[0000-0002-3514-0383]{G. Mark Voit}
\affiliation{Department of Physics and Astronomy, Michigan State University,
East Lansing, MI 48824, USA}
\email{}

\author[0000-0002-6413-4142]{Stephen A. Walker}
\affiliation{Department of Physics and Astronomy, University of Alabama in Huntsville,
301 Sparkman Drive, Huntsville, AL 35899, USA}
\email{}

% \author[orcid=0000-0000-0000-0001,sname='North America']{Tundra North America}
% \altaffiliation{Kitt Peak National Observatory}
% \affiliation{University of Saskatchewan}
% \email[show]{fakeemail1@google.com}

\vspace{5pt}
\begin{abstract}
We present JWST/NIRSpec IFU spectroscopy of the central $618\times618$ pc$^2$ ($\sim3''\times3''$) of NGC~4696, the BCG in the Centaurus cluster. Leveraging the $\sim0.1''$ ($20.6$ pc) pixel size of JWST, we resolve a compact circumnuclear rotating disk (radius of $\sim120$ pc) traced by \paa and \hs{1} emission, which allows a reassessment of the AGN position based on the kinematic centre of this disk. A central \paa velocity dispersion reaching $\sigma\sim449$ \kms implies a SMBH mass of $\sim10^9$ \msun, corresponding to a sphere of influence of $r_\mathrm{inf}\sim60$ pc, resolved by our observations. Position--velocity diagrams reveal an increase from $\sim-200$ to $\sim600$ \kms on scales of $\sim150$ pc (a gradient of $4.7$ \kms pc$^{-1}$) and an accretion rate of $\sim18$ \msunpyr feeding the CND. The \paa emission shows a double-component in the core, with a high-dispersion redshifted component reaching $\sigma\sim600$ \kms. In contrast, MUSE \ha observations covering the $\sim10$ kpc-scale filamentary structure recover only weak velocity gradients ($\lesssim150$ \kms) within $\sim300\times300$ pc$^2$ and do not resolve the disk due to larger PSFs and pixel sizes. ALMA CO(2--1) data reveal only compact molecular clumps within a $\sim410\times410$ pc$^2$ region, with no extended counterpart to the structures traced by \paa and \hs{1}. Stellar kinematics show a smooth velocity field and broad dispersion profile, clearly decoupled from both the multiphase gas and the hot ICM probed by XRISM. These results provide a direct, spatially resolved view of gas dynamics within the inner few hundred parsecs, demonstrating the power of JWST/NIRSpec to probe SMBH feeding.
\end{abstract}


\section{Introduction}

Supermassive black holes (SMBHs) at the centres of galaxies are key pieces in understanding the complex puzzle of galaxy growth and evolution. Through active galactic nuclei (AGN) feedback, the mechanical and radiative energy released by accretion onto the SMBH interacts with the surrounding medium, displacing hot interstellar gas from the galactic core and suppressing star formation \citep{fabian_observational_2012}. The optimal conditions to study AGN feedback are found in the largest known galaxies, which sit in the centre of all galaxy clusters. These brightest cluster galaxies (BCGs) offer an optimal environment for studying AGN feedback, as they reside in dense cluster cores where the hot intracluster medium (ICM) emits strongly in X-rays and typically peaks at the BCG location \citep[see][]{rhee_xray_1991,jones_structure_1984}. By tracing the displacement of gas via the formation of cavities, shock fronts and filaments, X-ray observations indicate that mechanical feedback from AGN radio jets is the dominant heating mechanism in cluster cores, nearly quenching star formation entirely \citep{mcnamara_mechanical_2012}. The resulting equilibrium between AGN-driven heating and radiative cooling is thought to regulate the thermodynamic state of cluster cores and to govern the complex gas dynamics observed around BCGs.

In an attempt to explain this presumed self-regulated loop between an SMBH and the surrounding ICM, various models have been proposed. \cite{pizzolato_nature_2005} suggest another model, in which the large mechanical energy injection induced by a major AGN outburst forms nonlinear density perturbations in the gas \citep[see also][]{soker_source_2006,pizzolato_solving_2010,revaz_formation_2008}. Some of these blobs cool rapidly as they fall towards the black hole, while others may form cold molecular clouds or stars. This model predicts that the optical filaments observed in multiple cooling flow clusters as well as the cool molecular gas traced by CO observations are mainly caused by the cooling ICM, in contrast to ``hot feedback" models that attribute these features to gas stripped from cluster galaxies \citep[e.g.][]{salome_cold_2003}. \cite{gaspari_chaotic_2013} expand this model by stating that frequent collisions allow even high angular momentum gas to get accreted by the black hole \citep[see also][]{Hlavacek-Larrondo2022,karenyang_how_2016,prasad_cool_2015}. Their model proposes that the frequent condensation of cold clouds and filaments leads to a cold, chaotic accretion flow whose dynamics are driven by stochastic, dissipative collisions that boost the accretion rate. \cite{mcnamara_mechanism_2016} rather propose that as radio AGN inflate rising X-ray bubbles, they simultaneously lift low entropy gas to altitudes where thermal instabilities can ensue and promote cooling in their wakes \citep[see also][]{revaz_formation_2008,li_simulating_2012}. These molecular clouds eventually return to the central galaxy to fuel star formation and the AGN. This mechanism, which they refer to as ``stimulated feedback" naturally sustains itself and is consistent with molecular clouds observed lying in the wakes of rising X-ray bubbles \citep{russell_alma_2017,olivares_ubiquitous_2019}. Together, these models emphasize that the gas flow in BCGs is governed by a proposed self-regulated cycle of heating and cooling, and they highlight the central role of thermal instabilities in forming the cold clouds that cool the ICM.

Another related model, the ``precipitation model", also invokes thermal instabilities to produce cold ($<100$ K) molecular clouds that feed the SMBH. However, its distinctive feature is the prediction that condensation occurs when the ratio $t_\mathrm{cool}/t_\mathrm{ff}$, the ratio of cooling time to free-fall time, drops below $10$ \citep{voit_regulation_2015,sharma_thermal_2012}. A key outcome of this model is that the large-scale gas infalling onto the SMBH follows chaotic turbulent motions \citep{gaspari_shaken_2018}.
% Although several observations have supported this type of chaotic accretion by detecting cold molecular gas through absorption line features against a BCG's bright continuum source \citep[e.g.][]{rose_constraining_2019}, this type of spectroscopy provides only line of sight velocities, limiting constraints on the full 3D kinematics of the accretion flow.
However, a challenge for the study of cold chaotic accretion is that it predicts that the condensation into cold clouds occurs within parsecs to a few hundred parsecs of the SMBH \citep{gaspari_shaken_2018}, which motivates the need for high spatial resolution observations.

\begin{figure*}
    \centering
    \includegraphics[width=\linewidth]{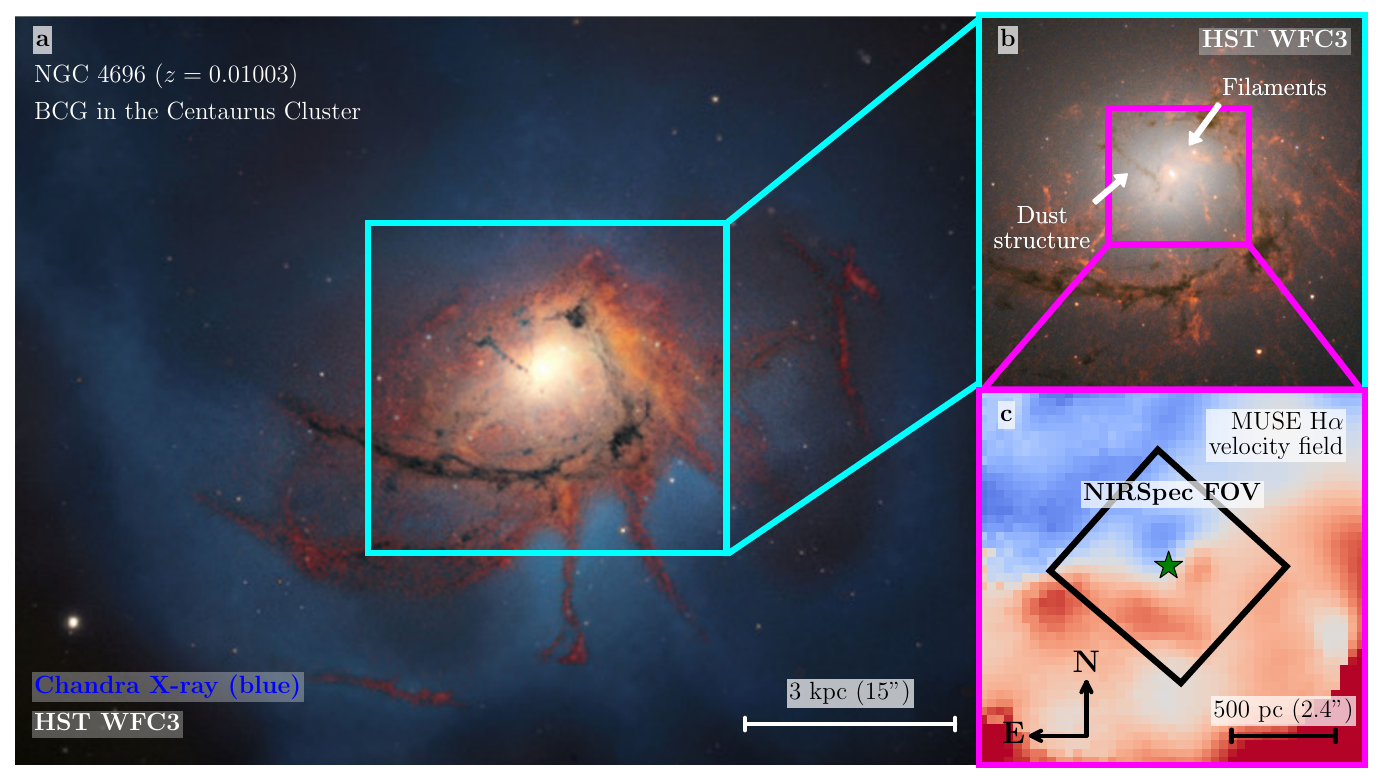}
    \caption{A multiwavelength view at NGC~4696. \textbf{a} The entire galaxy as seen by Chandra (blue) and by HST Wide Field Camera 3 (WFC3), which is a composite image of the F467M (blue), F814W (yellow) and F665N (red) filters. \textbf{b} Same HST image as panel \textbf{a} but zoomed on the core, which highlights a network of filaments and dust structures. \textbf{c} The MUSE \ha velocity field \citep{xrismcollaboration_bulk_2025,hamer_discovery_2019} in the central $\sim1.5\times1.5$ kpc$^2$ in which blueshifted (up to $\sim-100$ \kms) and redshifted (up to $\sim+100$ \kms) emission relative to the BCG is shown in blue and red respectively. The region targeted by our JWST/NIRSpec observations is overlayed. The green star is the AGN position aligned on HST \ha observations by \cite{fabian_hst_2016} based on VLBA observations by \cite{taylor_lowpower_2006} (see Section~\ref{sec:Black Hole Position}).}
    \label{fig:intro_fig}
\end{figure*}

To investigate the kinematics of the gas in the vicinity of a SMBH with unprecedented resolution, we target the bright elliptical galaxy NGC~4696 at the centre of the Centaurus cluster (see Fig.~\ref{fig:intro_fig}). At a redshift of $z=0.01003$ \citep{xrismcollaboration_bulk_2025} corresponding to a luminosity distance of 43.3 Mpc (206 pc/"), it is the second nearest BCG that allows to spatially resolve the gas within the Bondi radius, the first being M87. However, NGC~4696 features a much dimmer AGN than M87, which prevents the need for complex point spread function (PSF) modelling. NGC~4696 is embedded in an extensive multiphase nebula composed of cool filaments bright in optical emission lines, which share the same structure as the central dust lane and the X-ray gas \citep{crawford_extended_2005}.
% These filaments are often interpreted as structures formed in the wakes of buoyant gas bubbles from previous AGN outbursts.
High-resolution \textit{Hubble Space Telescope} (HST) imaging by \cite{fabian_hst_2016} showed that these filaments converge into a complex swirling structure around and within the Bondi radius of the SMBH.
% revealed their narrow widths ($\sim60$ pc) and showed that they converge into a complex swirling structure around and within the Bondi radius.
% They further suggested that magnetic fields were supporting the filaments by being in pressure balance with the surrounding hot gas. Additionally, Chandra X-ray observations identified a 0.5 keV gas component with a morphology closely matching the \ha filaments \citep{sanders_very_2016} and the soft X-ray emission around the filaments has been found to correlate to the \ha surface brightness and morphology \citep{olivares_haxray_2025}. Using Herschel data, \cite{mittal_herschel_2011} derived an unusually low star formation rate of $0.13$ \msunpyr, and showed that the [\ion{C}{2}] emission shares the same spatial distribution and kinematics as the \ha gas, suggesting a common heating mechanism. \cite{canning_detection_2011} detected $10^6$ K coronal emission in [\ion{Fe}{10}], the strength of which indicates that the emitting gas may be heated rather than condensing out of the ambient hot medium.
% Integral Field Unit (IFU) observations by \cite{farage_optical_2010} examined the kinematics of gas inflow along the brightest filament and found substantial amounts of dust and gas, which they interpreted as evidence for a minor-merger origin of the filaments \citep[see also][]{sparks_imaging_1989}. Further IFU observations by \cite{canning_deep_2011} also revealed broad and narrow velocity components to the emission-line gas with distinct morphologies.
This swirling structure inspired the kinematic study performed in this paper.

The nucleus of NGC~4696 hosts a low X-ray luminosity, radio-loud AGN \citep{taylor_magnetic_2002,dunn_particle_2004}, which drives the feedback cycle in the Centaurus cluster. The AGN is mechanically powerful \citep[cavity power of $P_\mathrm{cav}\sim10^{42}$ erg s$^{-1}$;][]{russell_radiative_2013}, launching radio jets that inflate a series of X-ray cavities in the surrounding ICM \citep{fabian_deep_2005,birzan_systematic_2004}.
% Deep Chandra imaging reveals multiple inner X-ray bubbles and outer semicircular edges in the hot gas, suggestive of episodic AGN activity over time \citep{fabian_deep_2005}. High-resolution Chandra observations further show shocks close to the nucleus and a weak 1.1--1.4 Mach number shock around the central cavities, indicating that the AGN maintains a tight balance between heating and cooling in the cluster core \citep{sanders_very_2016}.
Stellar velocity dispersion measurements in the galaxy of $\sigma_*\sim254$ \kms reveal that the stars have a velocity dispersion well above that of the gas \citep[$\sigma_\mathrm{gas}=127\pm10$ \kms;][]{olivares_ubiquitous_2019}.
% stars and $\sigma_\mathrm{gas}=127\pm10$ \kms using \ha reveal that the gas has a velocity dispersion well below that of the stars \citep{olivares_ubiquitous_2019}.
Similarly, \cite{xrismcollaboration_bulk_2025} measured a hot gas velocity dispersion of $\lesssim120$ \kms within the central $1'\times1'$, suggesting that the ICM is relatively quiescent and that the bulk of the jet's mechanical energy is not dissipated through turbulence.
% The low radiative efficiency of the AGN also indicates that it is accreting at a slow rate, despite the presence of substantial surrounding gas \citep{fabian_hst_2016}.

In light of this, NGC~4696 is a classic example of a cool core system containing multiphase gas in the form of filaments. Its proximity allows us to examine in detail the physical processes governing the heating and cooling of the ICM and to test models of AGN feeding. In this work, we employ \textit{James Webb Space Telescope} (JWST) Near-Infrared Spectrograph (NIRSpec) Integral Field Unit (IFU) observations to resolve the gas kinematics within the SMBH's sphere of influence (SOI), with particular attention given to the \ha spiral structure reported by \cite{fabian_hst_2016}. The objectives of this study are to characterize the kinematics and the velocity dispersion of both the ionized gas and the stars in the vicinity of the SMBH, using a combination of emission and absorption line diagnostics. We further search for evidence of a rotating gas disk close to the SMBH and assess the local stellar kinematics. Our analysis is complemented by archival observations from the Multi Unit Spectroscopic Explorer (MUSE) \cite[see][]{hamer_discovery_2019,xrismcollaboration_bulk_2025} as well as from the Atacama Large Millimeter/submillimeter Array (ALMA) (Canning et al. in preparation). In a first companion letter based on the same NIRSpec dataset, \cite{hlavacek-larrondo_jwst_2026} reveal a rotating multiphase circumnuclear disk (CND) kinematically connected to kiloparsec-scale filaments and reproduce the morphology and kinematics with magnetohydrodynamic simulations. In a second companion paper, O. Pereira et al. (submitted) provide a comprehensive study of all detected emission lines and perform an analysis of the molecular hydrogen lines in particular to understand underlying excitation mechanisms.

This paper is organized as follows. Observations and data reduction are presented in Section \ref{sec:Observations}. The observed kinematics of both gas and stellar components are presented and compared with other observations in Section \ref{sec:Results}. We discuss the kinematics of different regions, observed gaseous structures and black hole properties in Section \ref{sec:Discussion}, and our key results are summarized in Section \ref{sec:Conclusions}. Throughout this paper, we consider a $\Lambda$CDM cosmology with $H_0=70$ \kms Mpc$^{-1}$, $\Omega_m=0.3$ and $\Omega_\Lambda=0.7$ from the latest measurements by \cite{planckcollaboration_planck_2016}.

\section{Observations and Data Analysis}\label{sec:Observations}
\subsection{JWST NIRSpec/IFU}
\subsubsection{Data Collection and Data Reduction}\label{sec:Data Collection and Data Reduction}
The NIRSpec/IFU observations were taken on 5 June 2025 (Program ID 5354, PI: Hlavacek-Larrondo) using the F170LP/G235H filter-grating pair with a resolution of $R\sim2700$ (velocity resolution of $\sim110$ \kms), covering wavelengths from $1.64$ \um to $3.14$ \um in the rest frame of NGC~4696.
The observation used a medium cycling dithering pattern and each of the 10 exposures consisted in a single integration read out using 19 groups of the NRSIRS2 readout pattern, resulting in the total exposure of 7.7 hours.
The program also included a Leakcal exposure obtained at one of the dither positions using the same exposure configuration.
%to obtain a sufficient signal-to-noise ratio (SNR) for both gas emission lines and stellar continuum.
The dither pattern was employed to improve spatial sampling and suppress detector artifacts, including bad pixels and cosmic rays, with the LeakCal exposure used to characterize and remove stray light from the NIRSpec/MSA.

The data were reduced using JWST Build 12.2 with CRDS context 1464 pmap and standard STScI pipelines \citep{Bushouse2019}. The \texttt{snowblind}\footnote{\href{https://github.com/mpi-astronomy/snowblind}{https://github.com/mpi-astronomy/snowblind}} package replaced the default algorithm to mitigate snowball and shower artifacts, and \texttt{nsclean} was applied for $1/f$ noise subtraction \citep{Rauscher2024}. Exposures were drizzle-combined into a final cube sampled at 0.05'' per spaxel ($\sim$10 pc), with an additional outlier rejection step to remove any remaining bad pixels.
The spatially undersampled NIRSpec IFU data exhibit sinusoidal modulations (wiggles) in the spectra, arising from the undersampled PSF, which can significantly distort the continuum and emission line profiles \citep[e.g.,][]{Perna2023, Law2023}. We mitigated this effect using the \texttt{WICKED} package \citep{Dumont2025}, which models and removes the wiggle signal from each spectrum.
An additional cube was produced at 0.1'' per spaxel, matching the NIRSpec IFU slicer resolution, to achieve a higher signal-to-noise ratio; all results are derived from the 0.1'' cube unless otherwise stated.

The astrometry of the IFU can be uncertain given its small field of view (FOV), which does not contain any stars; we therefore assigned a new WCS to the cube by matching the spiral swirl observed in HST/WFC3 F665N \citep{fabian_hst_2016} to the morphology of the Pa$\alpha$ line map \citep{hlavacek-larrondo_jwst_2026}. The HST image was aligned with multi-visit mosaic processing \citep{2010bdrz.conf..382F}, which is part of the Hubble Advanced Products pipeline and allows to add multiple exposures and align them to \textit{Gaia}.

%The NIRSpec observations are taken in a field of view (FOV) of $\sim3''\times3''$ ($618\times618$ pc$^2$) with a pixel size of $0.1''$ ($20.6$ pc).
%Ten-position medium cycling dithers were employed to improve spatial sampling and mitigate artifacts such as bad pixels and cosmic rays, extending through dither overlap and drizzle-weighted reconstruction the pixel resolution of $0.05''$.
%Since these high-resolution present worse SNRs, results should be assumed to be from the $0.1''$ data unless otherwise mentioned.

%\textcolor{red}{Pre-fitting data reduction steps will be completed by @Joseph.}

\subsubsection{Emission Line Fitting}\label{sec:Emission Line Fitting}

%%% MAR 03/09/26 %%%

We model the spectrum of each spaxel using the Likelihood Optimization of gas Kinematics in IFUs (LOKI) software \citep[see, e.g.,][]{reefe_directly_2025}.  This is a flexible package for modeling IFU spectroscopy that allows the user to specify the model for the continuum and emission lines by toggling individual components. The spectral model we adopt can be represented as:
\begin{align}
    \begin{split}
        I_\nu(\lambda) &= E_*(\lambda)\bigg[\sum_iC_i(\lambda)\bigg(\sum_j w_jS_j(v) \otimes G_{*}(v)\bigg)\bigg] \\
        &+ E_{\rm gas}(\lambda)\bigg(\sum_{k}G_{k}(\lambda)\bigg)
    \end{split}
\end{align}
where $E_*$ and $E_{\rm gas}$ represent \citet{Calzetti2000} extinction laws, of the form $10^{-0.4E(B-V)k'(\lambda)}$, where we constrain $E(B-V)_{\rm *} = 0.44E(B-V)_{\rm gas}$.  $C_i$ are Chebyshev polynomials, which may be up to 10th order (excluding the 0th order / constant term), and which act multiplicatively on our stellar templates $S_j$.  These templates are convolved with a stellar line of sight velocity distribution (LOSVD) $G_*$, which is modeled as a Gaussian.  Emission lines are also modeled with Gaussian profiles $G_{k}(\lambda)$.

For our stellar templates, we use the BT-Settl library of stellar spectra \citep{Allard2012}, which are generated from the \textsc{Phoenix} model stellar atmosphere code \citep{Hauschildt1999}.  These models are optimized for the low surface temperature stars which dominate the NIR continuum emission of NGC~4696, and they include a physically motivated prescription for the formation of molecules and dust clouds in the outer atmosphere layers.  They are also incredibly high resolution, allowing us to compare them to the NIRSpec data at native sampling and get accurate stellar velocities.  We choose a grid of templates with effective temperatures $T_{\rm eff} \in (2000, 7000)~{\rm K}$, surface gravities $\log g \in (-0.5, +5.0)$, and metallicities $\log (Z/Z_\odot) \in (-0.5, +0.5)$.  The BT-Settl grids do not have full coverage in $\alpha$-enhancement, so we choose not to restrict this axis, allowing $[\alpha/{\rm Fe}] \in (-0.2, +0.6)$.  The NGC~4696 spectra show some absorption features in the stellar continuum at $\sim 1.77$ \um, $\sim 1.80$ \um, and $\sim 1.85$ \um~ which are not included in these models, so we mask out these regions to prevent biasing the stellar continuum models. These stellar absorption features may be absent from current stellar continuum models because they fall within a wavelength regime heavily affected by telluric absorption and therefore inaccessible to ground-based spectroscopic observations. Additionally, to facilitate the convolution with the LOSVD, the data are resampled onto a grid of constant $\Delta v\sim70$ \kms to allow for the exploitation of the convolution theorem \citep[as in][]{Cappellari2017}. Note that the stellar templates are only used for kinematic analysis and not for estimating other physical properties.

For the emission lines, line of sight velocities are limited to $\pm 800$ \kms, and full widths at half maximum (FWHMs) are limited to $\leqslant 1500$ \kms.  We do not tie any lines together kinematically, except for the fainter H$_2$ 1-0 $Q$ lines, which would otherwise be poorly constrained.  All line FWHMs are corrected for instrumental broadening, assuming a spectral resolution for NIRSpec's high-resolution gratings of $R = 2700$, corresponding to a FWHM velocity of $\sim 111$ \kms. As such, we cannot physically interpret velocity shifts below this value. LOKI can check for multiple velocity components for specific emission lines in each spectrum by performing a statistical F-test, which checks whether
\begin{equation}
    F=\frac{\left(\chi_A^2-\chi_B^2\right)/(p_B-p_A)}{\chi_B^2/(n-p_B)}
\end{equation}
exceeds a critical value in an $F$-distribution with $(p_B-p_A,n-p_B)$ degrees of freedom. In this equation, the $A$ and $B$ subscripts denote the models that were fitted respectively with $N$ and $N+1$ components, $\chi^2$ is the chi-squared statistic, $p$ is the number of model parameters and $n$ is the number of data points. The critical value in the $F$-distribution is chosen to correspond to a confidence threshold of 99.7\%, or $3\sigma$. When allowing to fit for more than one component, LOKI tests adding each component iteratively, starting with a single component and adding other components one at a time, and retains the highest number of components that passed the F-test\footnote{For more information, see the LOKI documentation: https://github.com/Michael-Reefe/Loki.jl.}.  All lines are modeled with a single velocity component, except \paa, which uses 2. The reasons behind this choice are discussed in Section~\ref{sec:Region Analysis}.

To prevent the multiplicative Chebyshev polynomials from being degenerate with the extinction curves, we perform an initial round of fits without any polynomials.  Then, for a second round, we fix the $E(B-V)$ parameters to the values found in this first round, and include the 10th order Chebyshev polynomials as corrective factors for the stellar continuum model. The output of this second round is used for studying the stellar kinematics (Section~\ref{sec:Kinematics of the Stars}).  Finally, in a third round, we swap out the physically motivated stellar continuum templates with a continuum template derived directly from the data, extracted from a background region which is observed to have no emission lines (see Fig.~\ref{fig:regions} \textbf{h}). This provides a more accurate fit to the continuum, thus giving us more reliable emission line maps (Section~\ref{sec:Gas Kinematics}).

In each round of fitting, LOKI determines the best parameters for this model in 2 steps.  In the first step, the emission lines are masked out, and a fit to the continuum is performed.  This fit is done with a Levenberg-Marquardt local least squares optimization algorithm, as implemented in \textsc{CMPFit} \citep{Markwardt2009}, and the uncertainties are calculated based on the covariance of the parameters during the fit.  Then, in the second step, the continuum model is subtracted, and a cubic spline is subtracted from the residuals (with lines still masked) to remove any continuum structure not captured in the model. Then the emission lines are fit to these residuals.  A cubic spline is used here, in addition to the actual continuum fit, to allow the emission lines to be accurately modeled, even in the case where the continuum model is over- or under-estimated near a line.  The line fit is performed using simulated annealing, a global optimization algorithm that is ideal for disentangling complex kinematic profiles of lines.

This procedure is applied for two configurations: the first is done over integrated spectra from several regions (see Section~\ref{sec:Region Analysis}) and the second is performed for every spaxel in the data cube (see Sections~\ref{sec:Gas Kinematics} and \ref{sec:Kinematics of the Stars}).

%%%%%%

\subsection{MUSE}

% MUSE wide-field mode (WFM) observations of NGC~4696 were originally presented by \cite{hamer_discovery_2019}. The dataset consists of multiple pointings with individual exposure times of 820 s, providing a combined FOV of approximately $70''\times70''$ ($\sim14\times14$ kpc$^2$). The observations cover the entire filamentary structure of NGC~4696, including the region probed by our NIRSpec IFU observations. The data span a wavelength range of $\sim4750$--9000~\AA\ and were obtained under seeing conditions of $\sim1.2''$, with a spatial sampling of $0.2''$ per spaxel.

% For this work, we use a reprocessed version of the data cube produced by Sun et al. (in preparation). The reduced cube was constructed from seven individual exposures using the MPDAF package \citep{2016ascl.soft11003B}, including mosaicking and variance propagation. Emission-line fitting was performed by tying the velocity offsets of the different transitions and modelling them as Gaussian functions, while the stellar continuum was modelled using pPXF \citep{cappellari_parametric_2004} together with E-MILES stellar templates \citep{vazdekis_uvextended_2016}, enabling the extraction of stellar velocity and velocity dispersion maps across the field.

The detail of the MUSE wide-field mode (WFM) observations and data analysis can be found in \cite{xrismcollaboration_bulk_2025}. We combined twenty-two MUSE observations from 2014 to 2023, with a total exposure time of 5.66 hours, under seeing conditions of 0.70$''$--1.21$''$ and airmass of 1.08--1.72. These observations provide a FOV of approximately $70''\times70''$ ($\sim14\times14$ kpc$^2$) which cover the entire filamentary structure of NGC~4696, including the region probed by our NIRSpec IFU observations.

Further sky subtraction was performed with the Zurich Atmosphere Purge software (ZAP). We fitted the combined data cube using the TARDIS data analysis pipeline\footnote{\url{https://gitlab.com/francbelf/ifu-pipeline}}, a Python-based spectral fitting program that incorporates the penalized pixel-fitting method \citep[pPXF; ][]{cappellari_parametric_2004}, based on Physics at High Angular resolution in Nearby GalaxieS (PHANGS)-MUSE Data Analysis Pipeline (DAP) \citep{emsellem_phangsmuse_2022}. The E-MILES stellar templates \citep{vazdekis_uvextended_2016} were used.

\subsection{ALMA}

ALMA band 6 $^{12}$CO J($2-1$) $\nu=0$ observations were taken as part of the Cycle 5 program 2016.1.01117.S (PI Canning). The C40-8 configuration was used enabling an angular resolution of 0.052'' and maximum recoverable scale of 0.57''. A bandwidth of 1875 MHz with 1.3 \kms resolution was used to trace velocity structure in the gas. The observations were reduced and data cubes obtained with \textsc{casa} version 5.1.1-5, software maintained by the National Radio Astronomy Observatory (NRAO) \citep{2007ASPC..376..127M}. The total exposure time was 11 hours. Continuum subtracted data cubes were created using \textsc{uvcontsub} and \textsc{clean}.

\section{Results}\label{sec:Results}

As found in \cite{hlavacek-larrondo_jwst_2026}, the JWST/NIRSpec IFU observations reveal a rotating multiphase CND connected to kiloparsec-scale filaments and reproduced with magnetohydrodynamic simulations. Here, we analyze the same JWST dataset using the LOKI fitting software to model the stellar continuum and simultaneously fit the full set of detected emission lines, extending beyond the \paa-focused analysis presented previously. Combined with archival MUSE and ALMA observations, this enables a more detailed investigation of both the gas and stellar kinematics within the NIRSpec FOV of NGC~4696, with particular emphasis on the \ha spiral structure reported by \citet{fabian_hst_2016}.

We first define a set of physically-motivated regions that are used throughout the analysis and examine their integrated spectra in Section~\ref{sec:Region Analysis}. We then investigate the dynamics of the gas in Section~\ref{sec:Gas Kinematics} and compare them with the stellar kinematics in Section~\ref{sec:Kinematics of the Stars}. Given the multiphase nature of the gas in NGC~4696, we also compare our JWST/NIRSpec results to complementary MUSE observations in Section~\ref{sec:MUSE Kinematics} as well as with ALMA data in Section~\ref{sec:ALMA Observations}.

We note that the kinematic centre of the rotating CND reported by \cite{hlavacek-larrondo_jwst_2026} appears to be offset from the AGN position reported by \cite{fabian_hst_2016} \citep[see also][]{taylor_lowpower_2006}. If the rotating CND is governed by the gravitational potential of the AGN, the AGN itself should coincide with the kinematic centre, corresponding to the location of zero rotational velocity. In this work, we therefore determine a new position for the AGN that coincides with the kinematic centre of the CND by modeling the velocity gradients across the CND. Details of this analysis are presented in Section~\ref{sec:Black Hole Position}. From this point onward, the new AGN position is marked by a red cross in all maps and likely provides a more robust estimate of the dynamical centre of NGC~4696.

\subsection{Region Analysis}\label{sec:Region Analysis}

\begin{figure*}
    \centering
    \includegraphics[width=\linewidth]{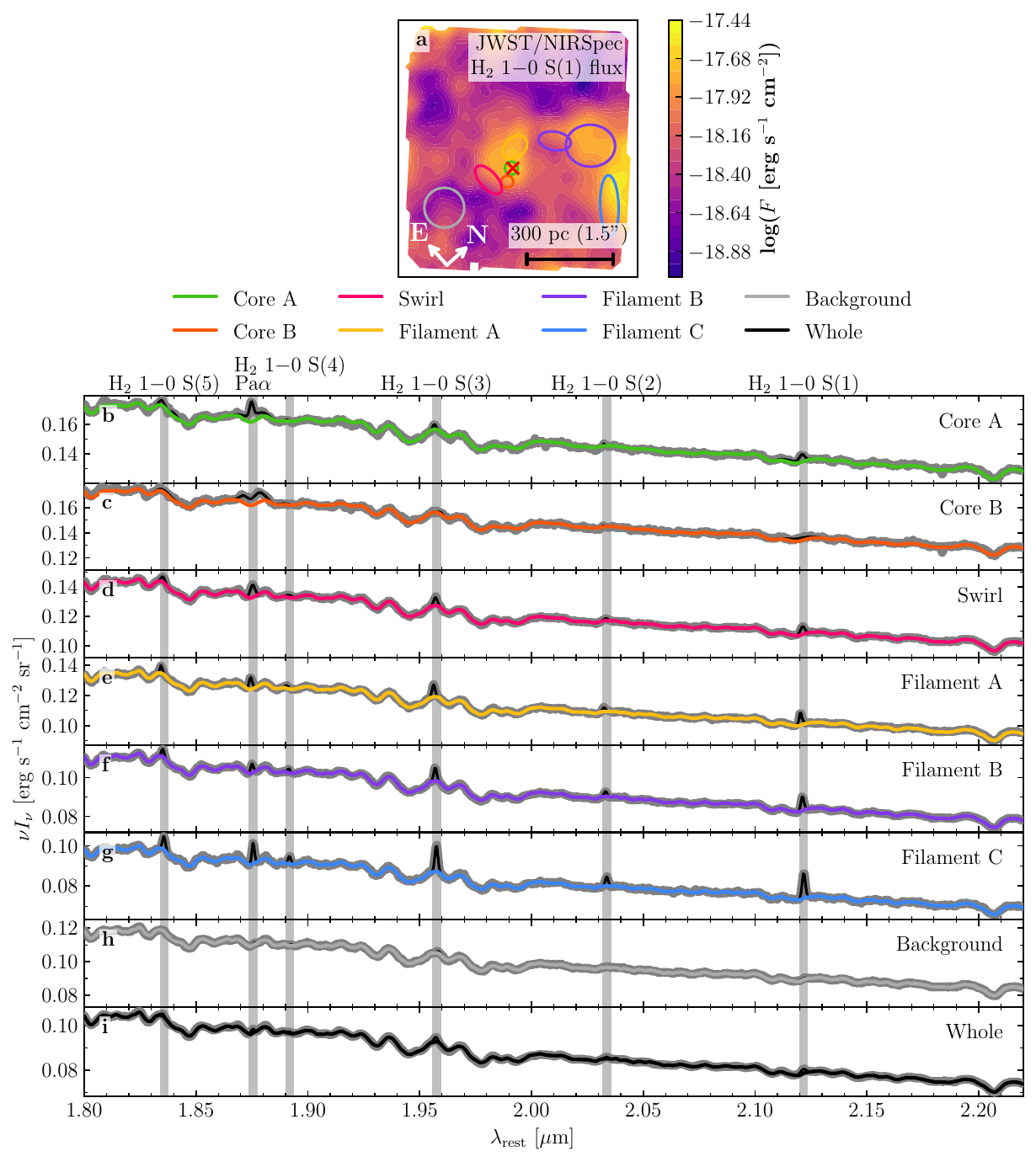}
    \caption{Regions established with EMPCA and their mean spectra useful for establishing the detectability of specific emission lines. \textbf{a} JWST/NIRSpec \hs{1} flux map fitted using LOKI and processed using the methods detailed in Section~\ref{sec:Gas Kinematics}, with regions defined using EMPCA (Section~\ref{sec:Region Analysis}). The red cross marks the kinematic centre (see Section~\ref{sec:Black Hole Position}). \textbf{b}--\textbf{i} Integrated spectra for each region in grey with their corresponding LOKI fits (continuum in color, emission lines in black) over the wavelength range encompassing \paa and the \hs{1} to S(5) transitions. In this high SNR regime, all S lines are included to assess their detectability in spaxel-level fits.}
    \label{fig:regions}
\end{figure*}

In this section, we create several regions and examine their integrated spectra in order to establish detection thresholds on several expected emission lines. We first performed a weighted expectation-maximization principal component analysis \citep[EMPCA;][]{bailey_principal_2012} to continuum-subtracted spectra of all spaxels having a signal-to-noise ratio (SNR) $\ge30$, isolated near each of the \paa, \hs{1}, and \hs{3} lines. The EMPCA was performed with 10 components, with the first four accounting for most of the relevant spectral features and the six others mostly accounting for irrelevant noisy features. As such, we used the space of only the first four components to cluster the spaxels with a Gaussian mixture model (GMM) having 10 components, enabling the creation of spectrally-coherent spatial groups. Overall, the EMPCA and GMM clustering allow to establish spatial regions in which all spaxels have similar line profiles and ratios. To refine the final apertures, we combined the EMPCA-clustered maps with physical features of note from HST \ha data \citep{fabian_hst_2016} and Very Long Baseline Array (VLBA) data \citep{taylor_lowpower_2006}. Certain regions were also established exclusively based on notable physical features. For an exhaustive discussion of this process, see O. Pereira et al. (submitted). The resulting regions are shown in Fig.~\ref{fig:regions} \textbf{a}. Due to the diamond-shaped footprint of the NIRSpec detectors, the FOV is rotated by $48^\circ$ clockwise, allowing the flux map to fill square subplots more efficiently. The \hs{1} flux map used as the background is described in Section~\ref{sec:Gas Kinematics}. Panels \textbf{b}--\textbf{i} present the integrated spectrum of each region along with the corresponding LOKI fit obtained using the procedure outlined in Section~\ref{sec:Emission Line Fitting}.

In total, we have 8 regions that correspond to different structures. The \textit{Core A} and \textit{Core B} regions represent a single group established from GMM clustering that was split into two in accordance with the radio peaks seen with VLBA \citep{taylor_lowpower_2006} and the bright spots seen in the HST \ha data \citep{fabian_hst_2016}, whereas the \textit{Swirl} region is established exclusively from the same feature seen with HST. The \textit{Filament A}, \textit{Filament B} and \textit{Filament C} regions correspond to the filaments seen with HST at the northwest (see Fig.~\ref{fig:intro_fig} \textbf{b}). Note that \textit{Filament B} is made from two ellipses which outline the same spatial group. The manually defined \textit{Background} region represents a region of low gas emission in all lines but which still contains a stellar population. The integrated spectrum of this region (Fig.~\ref{fig:regions} \textbf{h}) is the one used as a continuum model in the third and last round of fitting (see Section~\ref{sec:Emission Line Fitting}). Finally, the \textit{Whole} region corresponds to an aperture which integrates the entire field.

These integrated fits serve two main purposes. First, they provide robust constraints on the global kinematics of each gas component in a high SNR regime. Second, they establish which spectral features should be detectable in the lower SNR conditions encountered in individual spaxel fitting (Sections~\ref{sec:Gas Kinematics} and \ref{sec:Kinematics of the Stars}).

The \textit{Core A} and \textit{Core B} regions exhibit weaker molecular hydrogen emission than the other regions, in particular when compared to \textit{Filament C}'s strong (SNR $>25$) \hs{1} and S(3) emission. In contrast, \paa emission is detected in all regions  with SNR $>4$, except for the \textit{Background} region.

Even in the integrated spectra, we find no significant detections ($\mathrm{SNR} \lesssim 1$) of expected metal lines for AGN environments \citep[e.g.][]{rodriguez-ardila_nearinfrared_2011,lamperti_bat_2017,calabro_nearinfrared_2023}, including \ion{He}{1}, [\ion{Mg}{8}], [\ion{Al}{5}], [\ion{Al}{9}], [\ion{Si}{6}], [\ion{Si}{7}], [\ion{Si}{9}], [\ion{Si}{11}], [\ion{S}{11}], and [\ion{Fe}{2}]. These transitions are therefore excluded from both the regional and spaxel-level fitting procedures. Similarly, the Br$\delta$ and Br$\gamma$ lines are not detected ($\mathrm{SNR} \lesssim 1$) and are omitted from all fits.

The Br$\beta$ transition is blended with the \ho{2} emission line, with a rest-frame wavelength separation of $\sim 1$ nm. At the NIRSpec resolving power ($R \sim 2700$), this separation is comparable to the instrumental resolution ($\Delta\lambda \sim 1$ nm), preventing reliable decomposition of the two components. Including both transitions in the fit introduces degeneracies without providing additional physical constraints. We therefore model this blended feature with a single component, labelled as \ho{2}, while noting that a contribution from Br$\beta$ may be present.

We also searched for additional molecular hydrogen transitions, including \hs{0}, S(2), S(4), S(6), S(7), S(8), Q(6), Q(8), and O(4). These lines are only marginally detected ($\mathrm{SNR}\lesssim3$) in the integrated spectra. Consequently, they are retained in the regional fits (Fig.~\ref{fig:regions}) but excluded from the spaxel-level analysis.

Finally, the regional fits allow us to assess the presence of multiple kinematic components. Among all detected lines, only \paa exhibits statistically significant evidence for a double-component structure. We therefore restrict the multi-component analysis to \paa in the spaxel-by-spaxel fitting (Section~\ref{sec:Gas Kinematics}).

\subsection{Gas Kinematics}\label{sec:Gas Kinematics}

Next, the kinematic properties of the gas were derived by fitting the JWST/NIRSpec IFU data on a spaxel-by-spaxel basis using LOKI, following the methodology described in Section~\ref{sec:Emission Line Fitting}. The resulting flux, velocity \citep[relative to the stellar redshift of $z=0.01003$ within a $\sim1$ kpc radius; ][]{xrismcollaboration_bulk_2025}, and velocity dispersion maps (relative to the velocity centroids) for the four brightest emission lines (\paa, \hs{1}, S(3) and S(5)) are presented in Fig.~\ref{fig:gas_lines}, while additional lines are shown in Appendix~\ref{app:gas_lines}.

As mentioned in Section~\ref{sec:Region Analysis}, each map is rotated by $48^\circ$ clockwise since the position angle of the NIRSpec aperture is not aligned with the pixel axes. Consequently, cardinal directions are explicitly indicated in all figures. For display purposes only, the maps were convolved with a Gaussian kernel of one-pixel standard deviation and use contours selected to highlight large-scale structures while suppressing pixel-scale noise.

The \paa, \hs{1}, S(3), and S(5) emission lines all exhibit strong flux in the northwestern stellar cluster region (\textit{Filament C}) as well as in the vicinity of the kinematic centre (Fig.~\ref{fig:gas_lines}, left column). The corresponding velocity fields (centre column) reveal a complex kinematic structure: gas associated with the filament at the northeast is predominantly blueshifted, while gas on the opposite side of the nucleus to the west is redshifted. This pattern is most clearly observed in the \paa velocity field (panel \textbf{b}), which suggests rotation about an axis oriented approximately along the northeast--southwest direction, consistent to what is found by \cite{hlavacek-larrondo_jwst_2026}. The velocity dispersion maps (right column) generally peak near the kinematic centre. However, the exact location of the peak varies slightly between tracers (e.g., comparing \paa and \hs{1} in panels \textbf{c} and \textbf{f}), indicating that different gas phases may probe distinct dynamical components.

\begin{figure*}[h]
    \centering
    \includegraphics[width=0.75\linewidth]{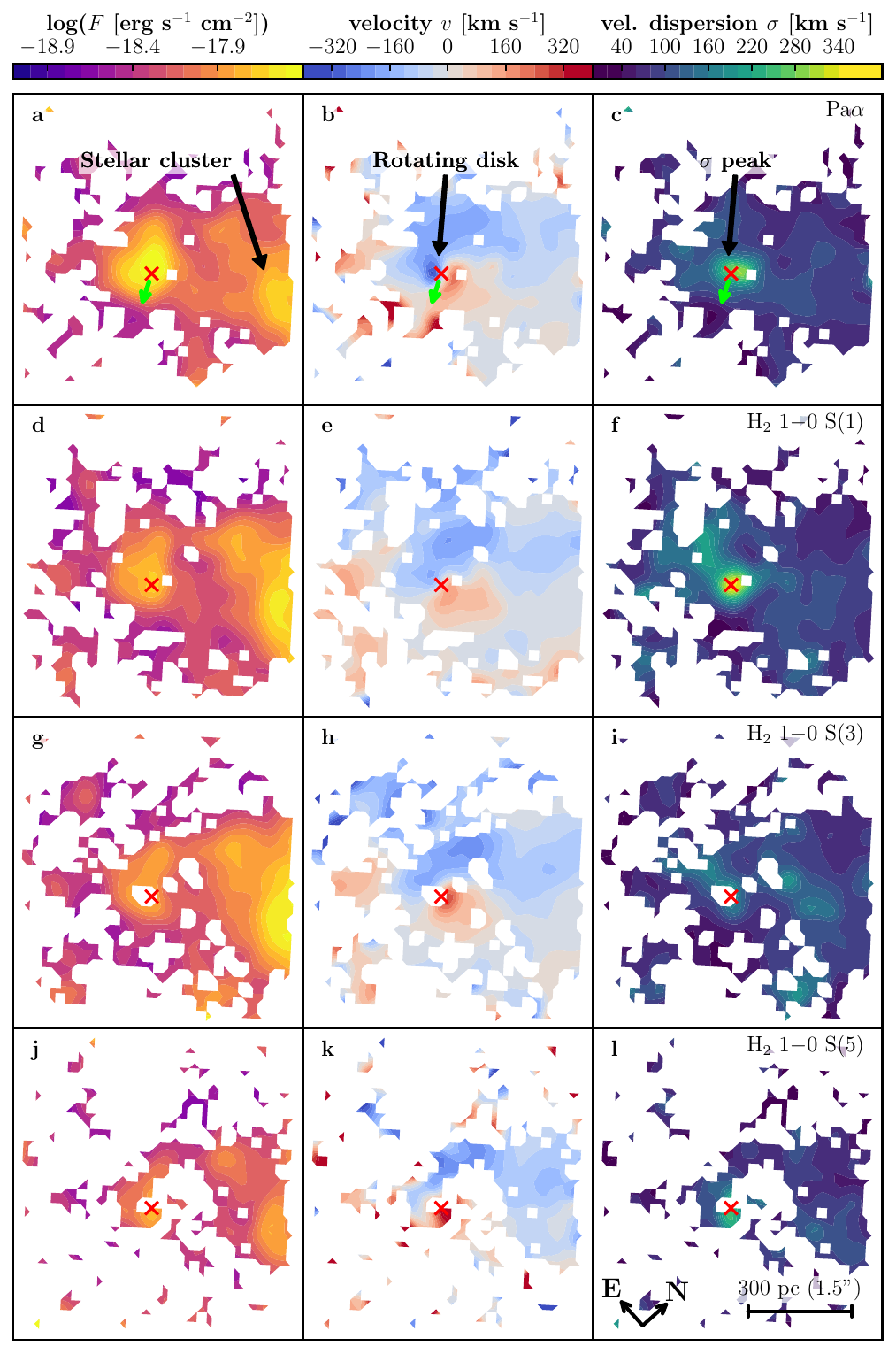}
    \caption{Gas kinematics derived from JWST/NIRSpec. Left to right: logarithmic flux, velocity, and velocity dispersion maps for the \paa, \hs{1}, S(3), and S(5) emission lines fitted using LOKI and processed using the methods detailed in Section~\ref{sec:Gas Kinematics}. All lines are fitted with a single component except \paa, for which only the blueshifted component is shown (see Fig.~\ref{fig:Pa_alpha_multi_component} for both \paa components). The red cross marks the position of the kinematic centre (see Section~\ref{sec:Black Hole Position}) and the green arrow in the top row represents the direction of the jet from VLBA observations \citep{taylor_lowpower_2006}. All shown emission has SNR $>2$.}
    \label{fig:gas_lines}
\end{figure*}

\begin{figure*}
    \centering
    \includegraphics[width=0.985\linewidth]{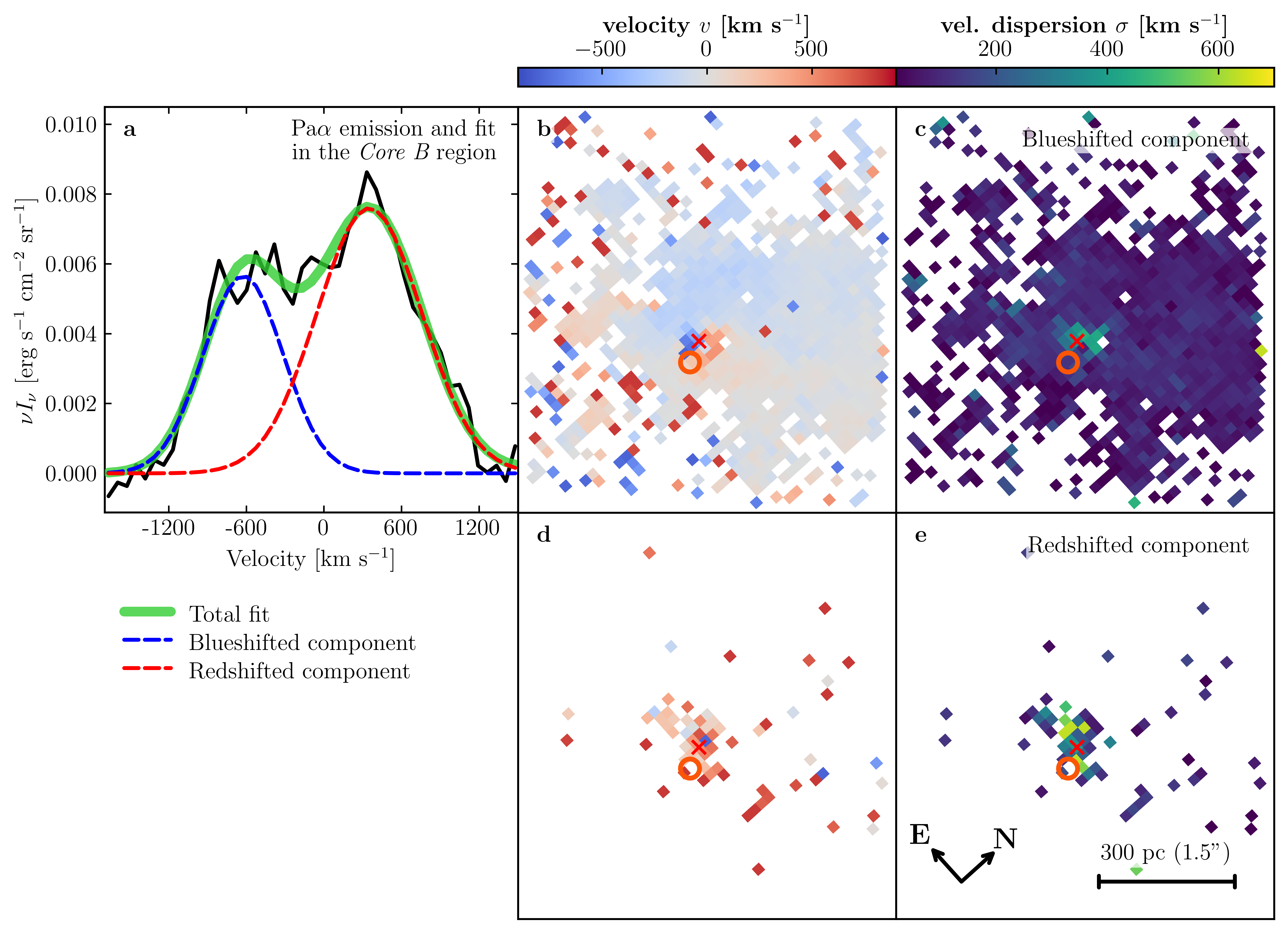}
    \caption{A closer view at the multi-component structure of the \paa emission, which highlights the need for a two-component fit. \textbf{a} The integrated spectrum around the \paa emission showing a bimodal distribution, taken from the \textit{Core B} region which is circled in orange in subplots \textbf{b}--\textbf{e}. The coloured curves show the result of fitting the data with LOKI. \textbf{b} and \textbf{c} The velocity and velocity dispersion fields of the blueshifted \paa component. \textbf{d} and \textbf{e} Same, but for the redshifted \paa component. Only spaxels with SNRs $>2$ are plotted. The data has been rotated as discussed in Section~\ref{sec:Gas Kinematics} but not smoothed in order to compare the two components directly.}
    \label{fig:Pa_alpha_multi_component}
\end{figure*}

As discussed in Section~\ref{sec:Region Analysis}, only the \paa emission line shows statistically significant evidence for multiple kinematic components. Each spaxel was therefore tested for the presence of up to two components using an F-test, as described in Section~\ref{sec:Emission Line Fitting}. The resulting double-component fits are shown in Fig.~\ref{fig:Pa_alpha_multi_component}. The integrated spectrum of the \textit{Core B} region (panel \textbf{a}) exhibits a non-Gaussian profile, motivating the use of multiple components. The spatial distribution of spaxels requiring two components (panel \textbf{b}) is concentrated primarily in a radius of $\sim50$ pc around the kinematic centre, with a smaller number of detections at larger radii. A total of 65 pixels are fitted with two components, which represents $\sim7.5\%$ of the total number of pixels with SNRs $>2$. The corresponding velocity dispersion map (panel \textbf{c}) shows the redshifted component typically exhibits a significantly higher dispersion than the blueshifted component, with typical values of $\sim370$ \kms within the central $\sim50$~pc, compared to $\sim250$ \kms for the blueshifted component.

Position--velocity (PV) diagrams were constructed to further investigate the gas kinematics along specific directions. These were generated using the \texttt{pvextractor} Python package \citep{ginsburg_radio_2015}, which extracts and combines spectra along user-defined apertures by averaging spaxels weighted by their spatial overlap. Typical applications of PV diagrams are performed on the raw spectra to perform an analysis independent of fitting \citep[for ALMA data, see e.g.][]{tremblay_galaxyscale_2018,oosterloo_closing_2024}. In our case, the strong and spatially varying stellar continuum in the NIRSpec data requires it to be subtracted by the continuum model obtained from LOKI prior to constructing the PV diagrams, isolating the emission-line signal. The noise level ($\sigma$) is estimated from the continuum-subtracted spectra in the 2.03--2.12 \um range (between \hs{2} and S(1)) and averaged over all non-edge spaxels. Detection thresholds (i.e. the contour levels) are then defined relative to this value.

\begin{figure*}
    \centering
    \includegraphics[width=\linewidth]{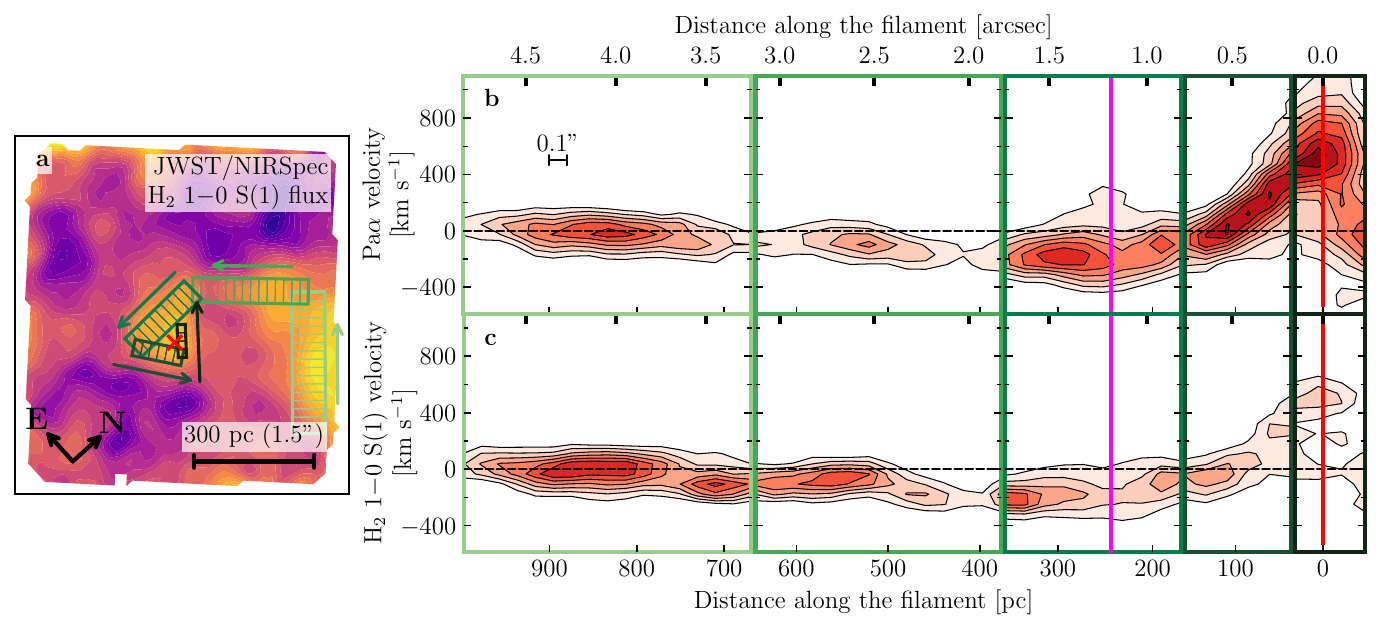}
    \caption{PV diagram of the gas flow along the filament, showing a velocity gradient and broadening along the last three apertures. \textbf{a} The \hs{1} flux map on which the PV apertures have been plotted. The first aperture is the light green one located in \textit{Filament C} at the northwest and the flow goes counter-clockwise following the arrows' directions. The background is the JWST/NIRSpec \hs{1} flux map fitted using LOKI and processed using the methods detailed in Section~\ref{sec:Gas Kinematics}. The red cross marks the position of the kinematic centre (see Section~\ref{sec:Black Hole Position}). \textbf{b} and \textbf{c} feature PV diagrams respectively for the \paa and \hs{1} emission lines, obtained using the stellar continuum subtracted data. Contours start at $3\sigma$ and increase in $2\sigma$ increments. The 0 position, highlighted by a vertical red line, corresponds to the position of the kinematic centre, which is in the last (black) aperture. The pixel size of $0.1''$ is also shown in panel \textbf{b}. The vertical magenta line roughly marks the transition region between the filament and the CND shown in \cite{hlavacek-larrondo_jwst_2026}.}
    \label{fig:PV_inflow}
\end{figure*}

Fig.~\ref{fig:PV_inflow} presents PV diagrams tracing the gas flow along the filament toward the kinematic centre. The apertures have been chosen to match the physically-motivated regions (see Fig.~\ref{fig:regions}), which are based on EMPCA as well as the HST \ha features \citep{fabian_hst_2016}. We also note that a second filament to the northwest of the kinematic centre can be observed (see Fig.~\ref{fig:gas_lines} \textbf{a}), but has been already studied using PV diagrams in \cite{hlavacek-larrondo_jwst_2026}.

In the first two apertures of Fig.~\ref{fig:PV_inflow}, corresponding respectively to \textit{Filament C} and \textit{Filament B}, both \paa and \hs{1} emission are centred near $0$ \kms. However, starting from the third aperture in \textit{Filament A}, an apparent velocity gradient emerges. The \paa emission (panel \textbf{b}) shifts from $\sim -200$ \kms to a broad distribution centred around $+600$ \kms in the innermost aperture at the kinematic centre position, indicating acceleration toward the nucleus. The \hs{1} emission (panel \textbf{c}) follows a similar trend but differs in its spatial and kinematic distribution: its flux is more concentrated in the filaments, while the emission near the nucleus is weaker compared to \paa. In any case, we still have detections above $3\sigma$, which come from extracting spectra in bins along the apertures, thus boosting the SNR. The acceleration of the gas is measured in Section~\ref{sec:Witnessing the Infall of Gas onto the SMBH}. Additionally, the \hs{1} profiles near the kinematic centre appear fragmented, exhibiting multiple peaks ($>3\sigma$) at distinct velocities rather than a single broad component. In Appendix~\ref{app:H2S1 Triple-Component Emission Profile}, we show a triple-peaked \hs{1} profile, taken just after the kinematic centre (see the last aperture in Fig.~\ref{fig:PV_inflow} \textbf{c}), which highlights the statistical significance of these components. Finally, we notice that after the kinematic centre, the velocity of the gas appears to return near 0 \kms, which actually highlights the gas feeding the CND through \textit{Filament A} (see in panel \textbf{a} the proximity between the last bin of the last aperture and the third aperture).

\begin{figure*}
    % Using the "interactive" environment creates a large blue box around the caption...
    % \begin{interactive}{animation}{pa_alpha_hires.mp4}
        \centering
        \includegraphics[width=\linewidth]{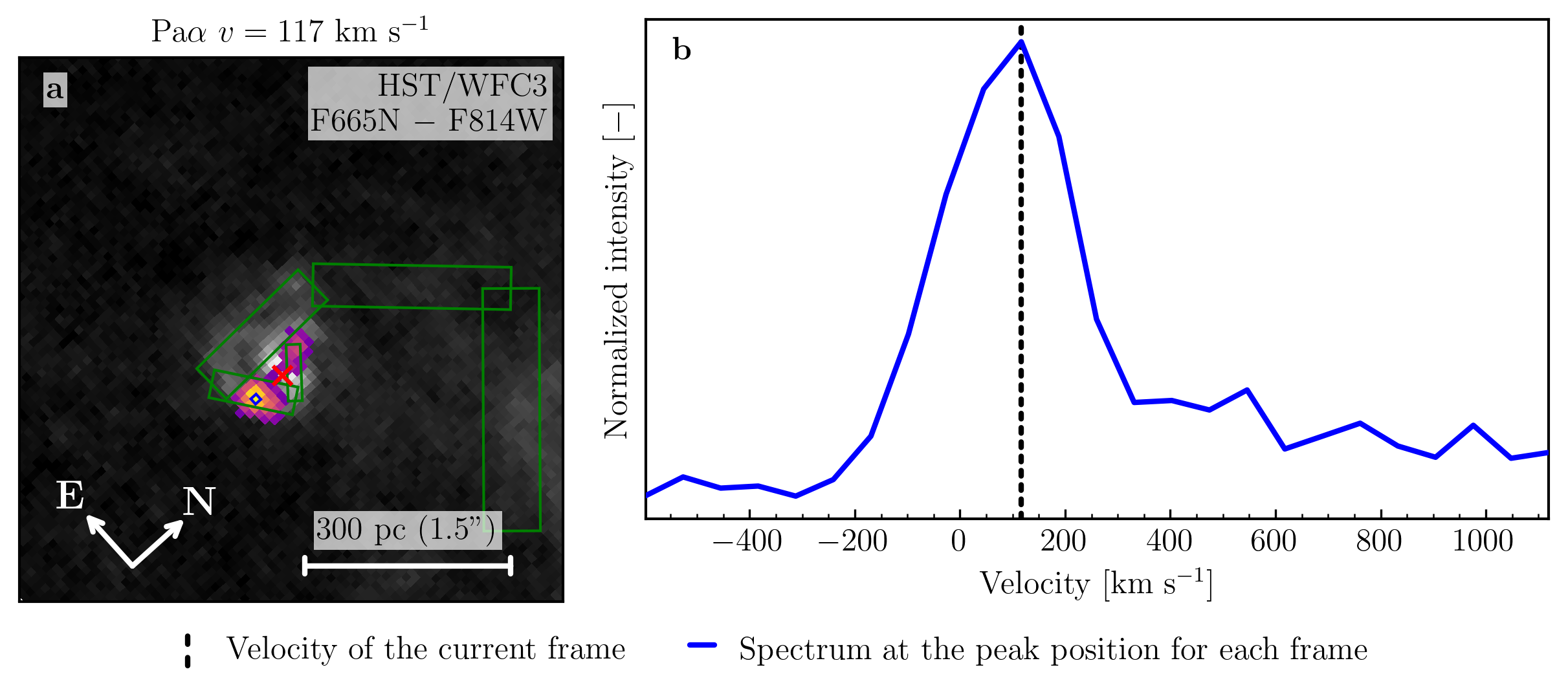}
    % \end{interactive}
    \caption{\paa flux at different velocities, showing the velocity gradient along the \textit{Swirl}. This figure is available as an animation in the HTML version of this paper as well as channel maps in Appendix~\ref{app:Pa alpha Channel Maps}. The movie is five seconds long and shows how the \paa emission peak shifts from $-600$ \kms at the start to $+1100$ \kms at the end along the HST swirl in a counter-clockwise motion, terminating at the location of the kinematic centre. \textbf{a} The high resolution (pixel size of $0.05''$) continuum subtracted data plotted at different wavelengths which map the \paa emission as a function of velocity. A threshold was applied to remove insignificant emission. The background image is the HST F665N \ha image subtracted by the F814W filter and the green apertures represent the suggested gas channels (same as in Fig~\ref{fig:PV_inflow}). The red cross marks the position of the kinematic centre (see Section~\ref{sec:Black Hole Position}). For each velocity the blue box highlights the peak spaxel, whose spectrum is shown in panel \textbf{b}. The moving vertical dotted line shows the current slice's velocity.}
    \label{fig:Pa_alpha_gif}
\end{figure*}

The kinematic evolution of the flowing gas is further illustrated in Fig.~\ref{fig:Pa_alpha_gif}, which shows the continuum-subtracted \paa intensity as a function of velocity. This figure is an animation and should be viewed on the html version of this paper. A static version is also available in Appendix~\ref{app:Pa alpha Channel Maps}. In panel \textbf{a}, we plot the stellar continuum subtracted cube emission at different velocities of the \paa emission line. In panel \textbf{b} we see the spectrum of the spaxel with the brightest emission at each specific velocity, allowing us to visualize the shift in velocity in velocity-space. The peak emission shifts continuously from the filament toward the kinematic centre following the swirl with increasing velocity, reaching $\sim1000$ \kms in the nuclear region. Similar trends, though less pronounced, are observed in other tracers such as \hs{1} and S(3).

\begin{figure}
    \centering
    \includegraphics[width=0.994\linewidth]{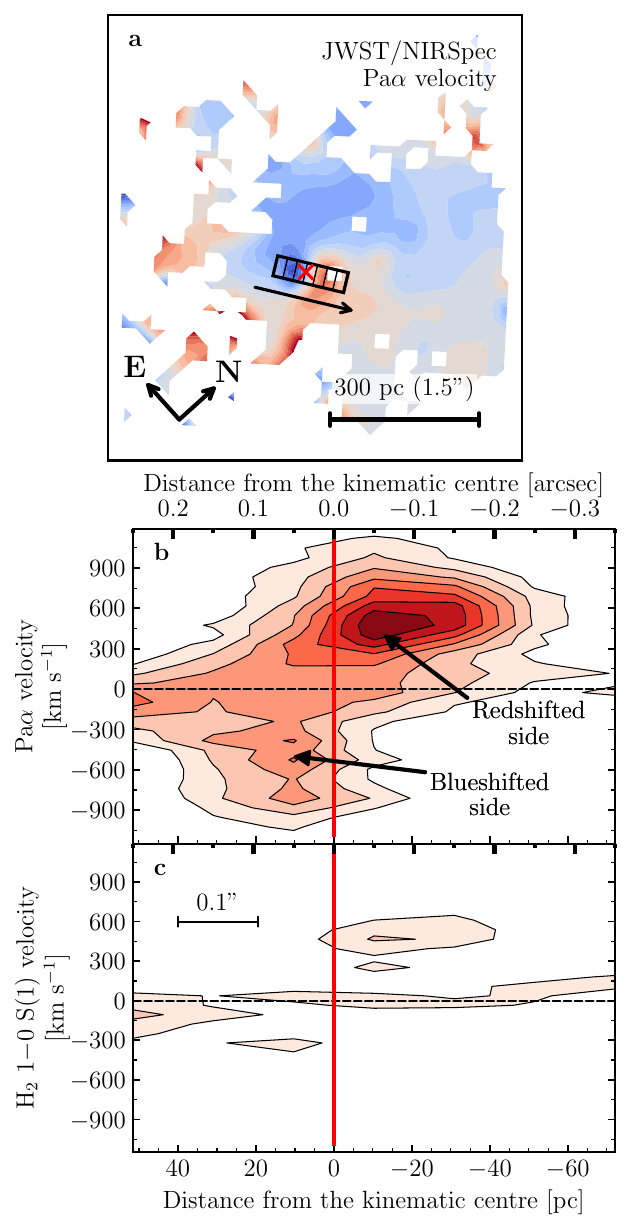}
    \caption{PV diagram of the gas in the plane of the CND, which highlights the blue- and redshifted sides. \textbf{a} The \paa velocity map on which the PV aperture has been plotted. The arrow indicates the direction of decreasing distance from the kinematic centre. The background is the JWST/NIRSpec \paa velocity map fitted using LOKI and processed using the methods detailed in Section~\ref{sec:Gas Kinematics} and filtered with SNR $>2$. The red cross marks the position of the kinematic centre (see Section~\ref{sec:Black Hole Position}). \textbf{b} and \textbf{c} feature PV diagrams respectively for the \paa and \hs{1} emission lines, obtained using the stellar continuum subtracted data. Contours start at $3\sigma$ and increase in $2\sigma$ increments. The 0 position, highlighted by a vertical red line, corresponds to the position of the kinematic centre and positions after this are negative by convention. The pixel size of $0.1''$ is also shown in panel \textbf{c}.}
    \label{fig:PV_AGN}
\end{figure}

Additional PV diagrams extracted along an aperture aligned with the kinematic position angle of the disk are shown in Fig.~\ref{fig:PV_AGN}. The \paa emission (panel \textbf{b}) exhibits significant line broadening around the kinematic centre and highlights the presence of a disk, with a blueshifted side reaching $\sim-1000$ \kms broadly distributed $\sim10$ pc from the kinematic centre and a redshifted side reaching $\sim1100$ \kms also $\sim10$ pc from the kinematic centre. In contrast, the \hs{1} emission (panel \textbf{c}) hints at the presence of a similar blue-/redshifted structure at similar scales but does not display a smooth or significantly broadened distribution as \paa.

\subsection{Kinematics of the Stars}\label{sec:Kinematics of the Stars}

\begin{figure}
    \centering
    \includegraphics[width=\linewidth]{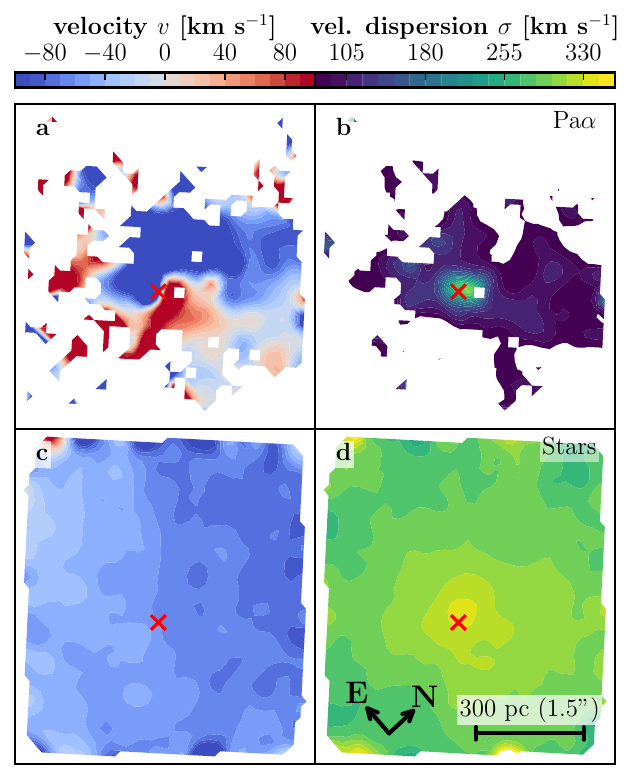}
    \caption{Stellar kinematics obtained by fitting stellar templates to the JWST/NIRSpec data using LOKI. Panels \textbf{a} and \textbf{b} show the \paa velocity and velocity dispersion maps as shown in Fig.~\ref{fig:gas_lines} only for comparison. As such, colorbars are chosen to be shared within columns. \textbf{c} The stellar velocity field obtained with LOKI. A small velocity gradient is observed but corresponds to a smaller shift than the NIRSpec velocity resolution at $R\sim2700$ ($\sim110$ \kms) and could occur from known problems with the JWST/NIRSpec IFU calibration distortion model\footnote{\url{https://www.stsci.edu/files/live/sites/www/files/home/jwst/documentation/technical-documents/\_documents/JWST-STScI-009029.pdf}}. \textbf{d} The stellar velocity dispersion map showing a gradual increase in velocity up to the kinematic centre. All maps have been processed using the methods detailed in Section~\ref{sec:Gas Kinematics}. The red cross marks the position of the kinematic centre (see Section~\ref{sec:Black Hole Position}).}
    \label{fig:stellar_kinematics}
\end{figure}

The stellar kinematics were derived from the second stage of the fitting procedure described in Section~\ref{sec:Emission Line Fitting}, during which stellar templates are fit prior to the final continuum substitution used to optimize the emission-line measurements. These fits are mainly driven by the CO $\Delta v=2$ band heads and provide spatially resolved estimates of the stellar velocity and velocity dispersion, shown in the bottom row of Fig.~\ref{fig:stellar_kinematics}. However, they do not provide other physical properties such as age or mass since our fitting routine fits a collection of individual star templates whilst allowing their weights to be free (see Section~\ref{sec:Emission Line Fitting}). For comparison, the corresponding \paa kinematics are reproduced in the top row of Fig.~\ref{fig:stellar_kinematics}, with identical color scales applied within each column, which highlight extensive differences between the behaviours of the gas and the stars.

\begin{figure*}
    \centering
    \includegraphics[width=0.65\linewidth]{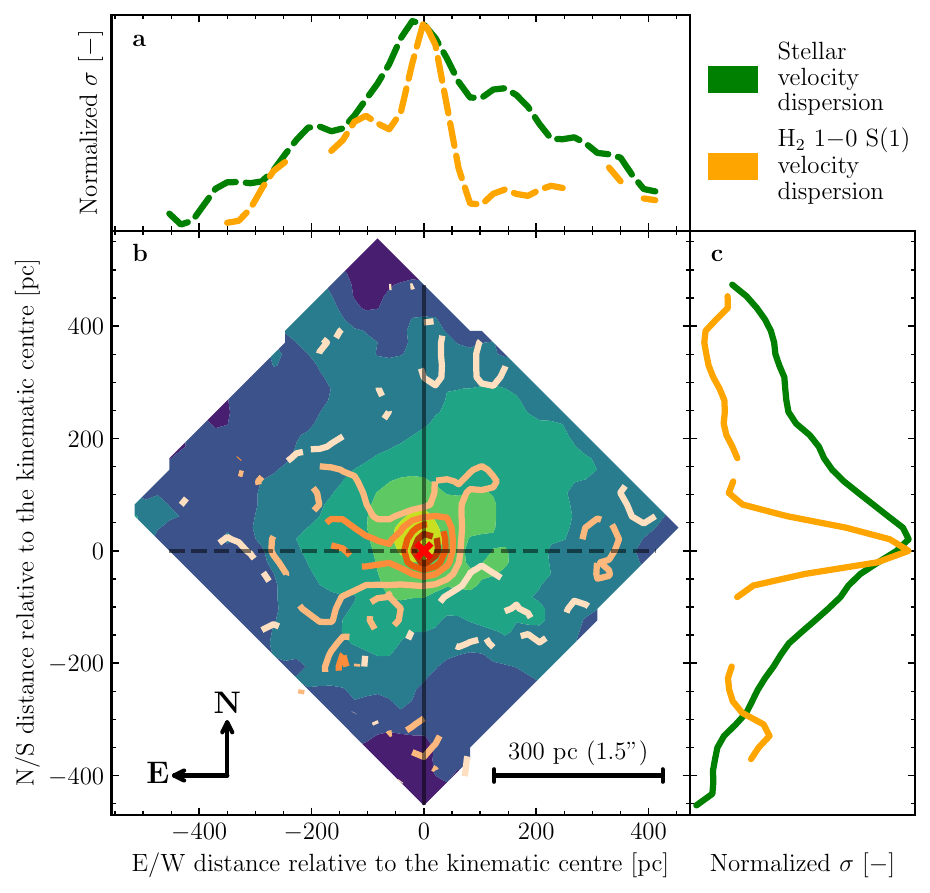}
    \caption{Velocity dispersion comparison between the stars and the gas, highlighting the broader radial profile of the stellar velocity dispersion compared to the gas. \textbf{b} The background image (purple to green) shows the normalized stellar velocity dispersion map and the overplotted unfilled orange contours represent the normalized \hs{1} velocity dispersion map. The \hs{1} map has been filtered with an SNR cut of 2, which introduces Not a Number (NaN) values that cause discontinuous lines in all subplots. For both maps, the data has been cropped at the edges to remove unphysical values present at the edges and it has been smoothed with a gaussian kernel having a standard deviation of 1 pixel. The red cross is the position of the kinematic centre. The dashed horizontal line and the solid vertical lines represent the projections along which the velocity dispersions shown respectively in panels \textbf{a} and \textbf{c} have been sampled. These allow to compare the distributions of the gas and of the stars along different axes. Note that this figure shows the unrotated JWST/NIRSpec data since the projections are made along the natural axes of the pixels.}
    \label{fig:sigma_comparison}
\end{figure*}

A more direct comparison between the stellar and gas velocity dispersions is presented in Fig.~\ref{fig:sigma_comparison}. The peaks of the stellar and gas velocity dispersion distributions are spatially coincident within uncertainties, indicating both components respond to the same central potential. However, the radial profiles differ markedly: the stellar velocity dispersion exhibits a broad, smoothly varying profile, whereas the gas velocity dispersion is sharply peaked and declines more rapidly with distance from the centre. This difference further supports the interpretation that the stellar component traces the global gravitational potential, while the gas is subject to additional localized dynamical processes.

\begin{figure*}
    \centering
    \includegraphics[width=\linewidth]{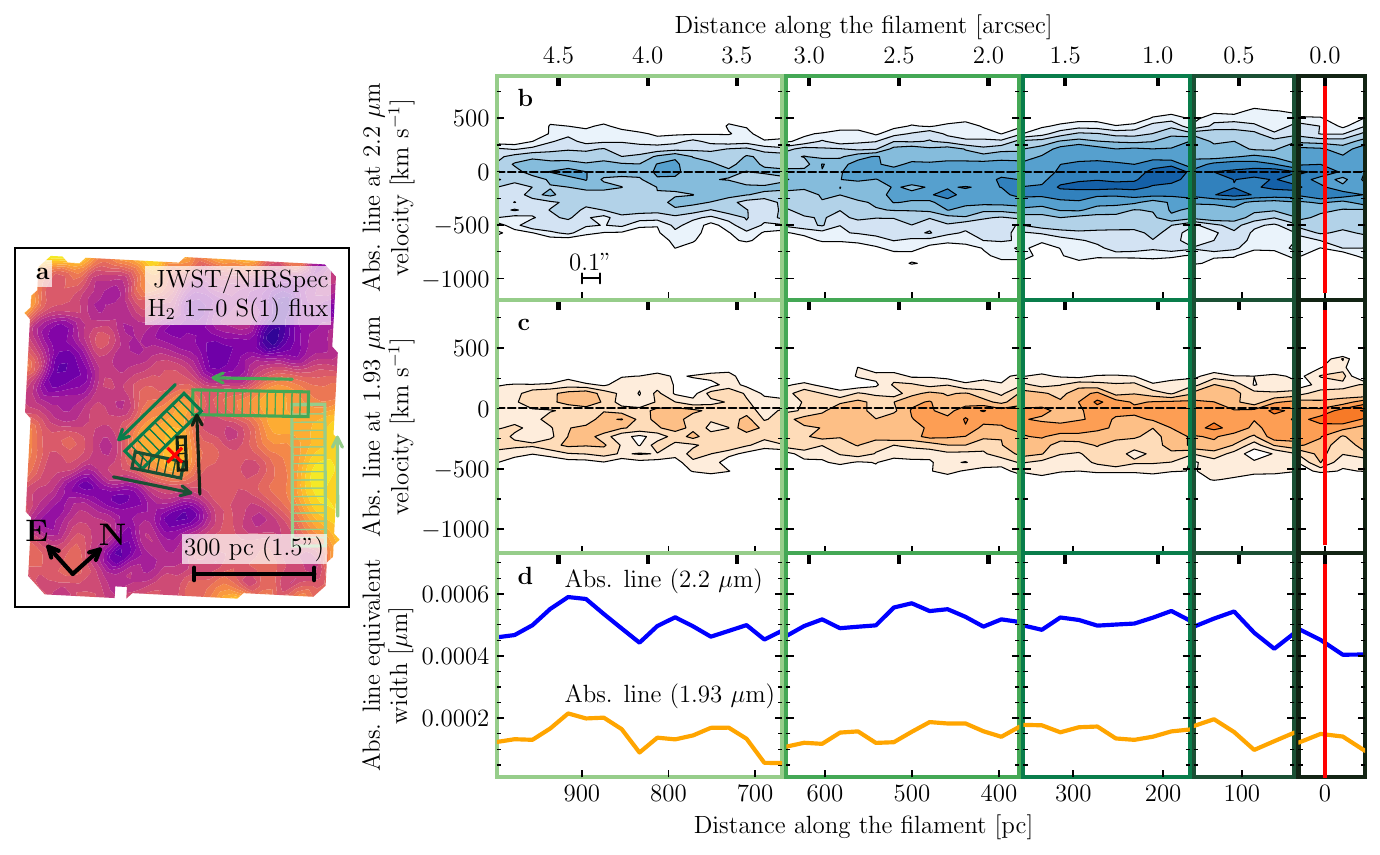}
    \caption{PV diagram of two stellar absorption line features at 2.2 \um and 1.93 \um along the filament and their equivalent widths, showing no velocity shift, an approximately constant line width contrarily to the gas PV diagrams (Figs~\ref{fig:PV_inflow} and \ref{fig:PV_AGN}) and no increase in equivalent width. \textbf{a} The \hs{1} flux map on which the PV apertures have been plotted. The first aperture is the lightest one located in \textit{Filament C} at the northwest and the flow goes counter-clockwise following the arrows' directions. The background is the JWST/NIRSpec \hs{1} flux map fitted using LOKI and processed using the methods detailed in Section~\ref{sec:Gas Kinematics}. The red cross marks the position of the kinematic centre (see Section~\ref{sec:Black Hole Position}). \textbf{b} and \textbf{c} feature PV diagrams respectively for the 2.2 \um and 1.93 \um absorption features, obtained by subtracting the raw data by a local linear fit around each absorption line. Contours mark the depth of the absorption features, start at $3\sigma$ and increase in $2\sigma$ increments. The 0 position, highlighted by a vertical red line, corresponds to the position of the kinematic centre of the gas, which is in the last (black) aperture. The pixel size of $0.1''$ is also shown in panel \textbf{b}. \textbf{d} The equivalent width of the 2.2 \um (blue) and 1.93 \um (orange) absorption lines along the filament.}
    \label{fig:PV_absorption_inflow}
\end{figure*}

To further investigate this and the stellar kinematics, we construct other PV diagrams in Fig.~\ref{fig:PV_absorption_inflow} along the same apertures as those used for studying gas flow (Fig.~\ref{fig:PV_inflow}) using the two deepest absorption features at 2.2 \um and 1.93 \um, which are documented absorption lines in stellar absorption models. We also plot the equivalent width of both absorption lines along these apertures, calculated using the merged spectrum of each bin along each aperture. Note that we do not use the CO band heads as these sometimes fall within the chip gap region throughout the field (i.e. the physical gap between the NIRSpec detectors that prevents collecting spectra in variable wavelength ranges). Contrary to the gas PV diagrams, these were generated from the raw data by fitting and subtracting a local \textit{linear} continuum around each feature, selected in emission line-free spectral regions. The absorption lines were specifically selected due to their location in marginally flat portions of the continuum. Velocity calibration was performed by identifying the rest wavelength of the same absorption features in a reference spectrum from a cold star \citep[LHS 1140, M dwarf star spectrum from][]{cadieux_transmission_2024}.

The resulting PV diagrams show an increase in absorption depth from the filamentary regions toward the kinematic centre, rising from $\sim 6\sigma$ to $\sim 9\sigma$ in panel \textbf{b} and from $\sim5\sigma$ to $\sim7\sigma$ in panel \textbf{c}. In contrast to the emission-line PV diagrams (Fig.~\ref{fig:PV_inflow}), the absorption features do not exhibit velocity shifts along the apertures, and their equivalent widths remain approximately constant.

\subsection{MUSE Kinematics}\label{sec:MUSE Kinematics}

\begin{figure*}
    \centering
    \includegraphics[width=\linewidth]{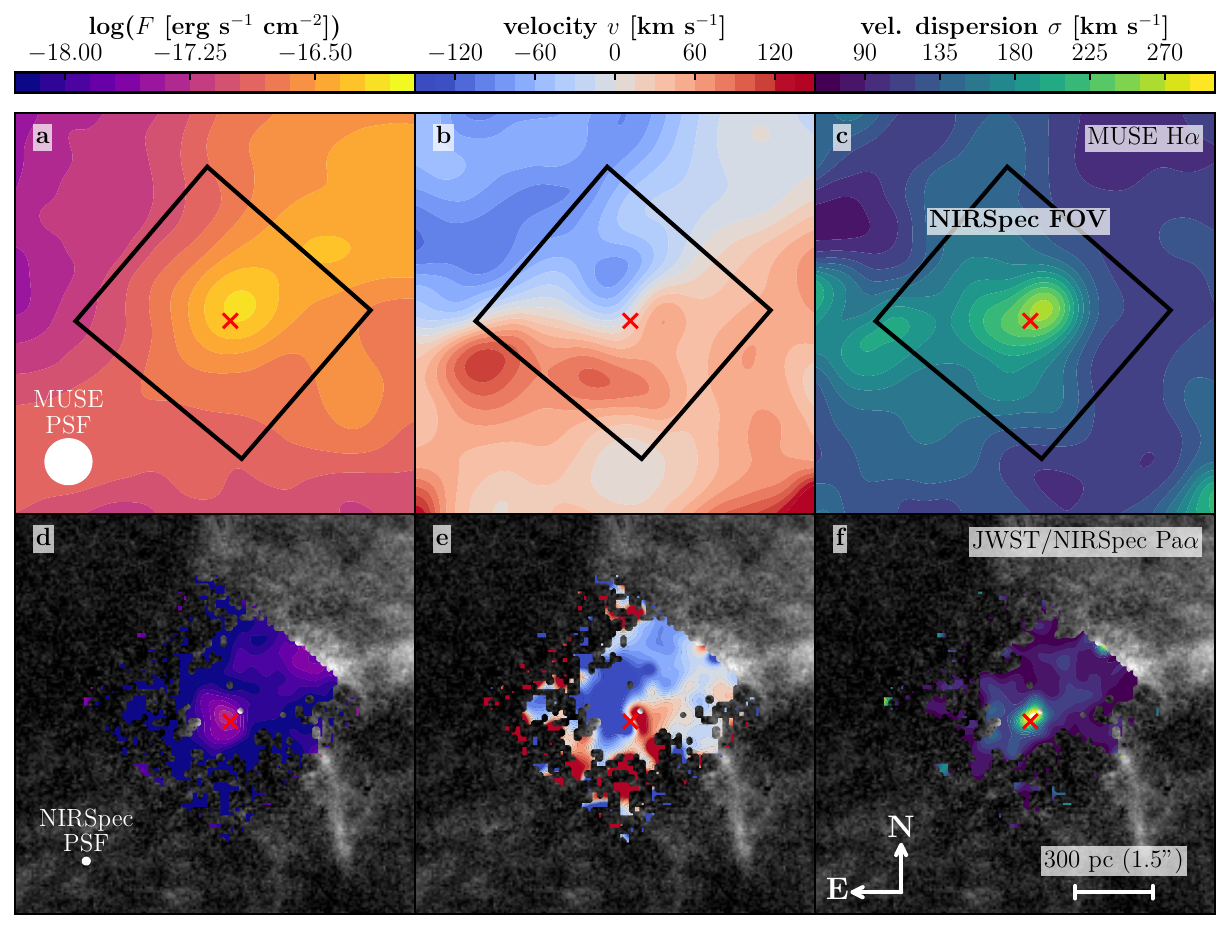}
    \caption{\ha kinematics derived from MUSE compared to \paa from JWST/NIRSpec. Left to right: logarithmic flux, velocity, and velocity dispersion maps, with colormaps shared within columns. The MUSE observations shown in the top row were convolved with a gaussian kernel having a standard deviation of one pixel and are plotted using contours. The NIRSpec FOV is also shown with the black square on top of the MUSE \ha data. In the bottom row, the HST F665N \ha image subtracted by the F814W filter is used as a background for the \paa maps, which were smoothed using the methods detailed in Section~\ref{sec:Gas Kinematics}. The \paa pixels have SNRs $>2$. The red cross marks the position of the kinematic centre (see Section~\ref{sec:Black Hole Position}). The PSF FWHM of both instruments is also plotted, with MUSE being seeing-limited ($\sim0.90''$) and NIRSpec having a PSF FWHM of $\sim0.17''$ at \paa's wavelength \citep{jones_blackthunder_2026}.}
    \label{fig:muse_gas_lines}
\end{figure*}

A comparison with the MUSE observations of NGC~4696 \citep[see][]{hamer_discovery_2019,xrismcollaboration_bulk_2025} provides a complementary, multiwavelength view of the gas and stellar kinematics on larger spatial scales ($\sim10$ kpc). Given the MUSE WFM resolving power at \ha of $R\sim2500$,\footnote{\url{https://www.eso.org/sci/facilities/paranal/instruments/muse/inst.html}} this corresponds to a spectral resolution of $\sim120$ \kms, very similar to NIRSpec's $\sim110$ \kms resolution. Fig.~\ref{fig:muse_gas_lines} presents the \ha flux, velocity, and velocity dispersion maps over the central $\sim1.6\times1.6$ kpc$^2$ ($\sim7.6''\times7.6''$), with the NIRSpec FOV overlaid. We did not perform a detailed modelling of MUSE's PSF and therefore plot a rough estimate based on the seeing. We also use a PSF of $\sim0.17''$ for NIRSpec at the wavelength of \paa \citep{jones_blackthunder_2026}.

The \ha flux distribution (panel \textbf{a}) broadly resembles that of \paa (panel \textbf{d}), with both tracers showing enhanced emission near the kinematic centre and along the filament connecting the northwestern stellar cluster to the kinematic centre.
% However, the velocity field derived from \ha (panel \textbf{b}) does not exhibit the same level of complexity as that observed in \paa (panel \textbf{e}). In particular, the pronounced redshifted and blueshifted structures of the CND detected in the NIRSpec data on opposite sides of the kinematic centre are not recovered in the MUSE observations. Similarly, the \ha velocity dispersion map (panel \textbf{c}) displays a centrally peaked distribution reaching only $\sigma_\mathrm{MUSE}\sim280$ \kms compared to JWST (panel \textbf{f}), which reaches $\sigma_\mathrm{JWST}\sim450$ \kms in \paa. These differences likely reflect the lower spatial resolution of MUSE, which is further compared to NIRSpec's in Section~\ref{sec:Comparison to MUSE}.
The velocity field (panel \textbf{b}) displays a global rotation around the kinematic centre whereas the velocity dispersion map (panel \textbf{c}) displays a centrally peaked distribution reaching $\sigma\sim280$ \kms.
% The differences between the MUSE and JWST/NIRSpec results caused by different PSFs is further studied in Section~\ref{sec:Comparison to MUSE}.}

\begin{figure}
    \centering
    \includegraphics[width=\linewidth]{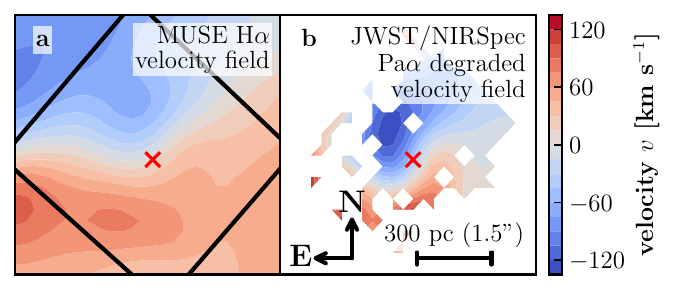}
    \caption{Comparison between the MUSE \ha velocity field and a degraded JWST/NIRSpec \paa velocity field post-processed to match MUSE's conditions. \textbf{a} The MUSE \ha velocity field, as plotted in Fig.~\ref{fig:muse_gas_lines}. The NIRSpec FOV is plotted with the black square. \textbf{b} The JWST/NIRSpec \paa velocity field from Fig.~\ref{fig:gas_lines} \textbf{b}, which has been filtered with an SNR threshold of 2, convolved with a gaussian having a FWHM equal to the MUSE observations' seeing ($0.90''$) and then reprojected to the spatial sampling of MUSE (pixels two times bigger than NIRSpec's). We do not see the CND and the kinematics now much more resemble the \ha MUSE observations. The red cross marks the position of the kinematic centre (see Section~\ref{sec:Black Hole Position}).}
    \label{fig:jwst_binning}
\end{figure}

The absence of the CND in the \ha MUSE data is coherent once the difference in PSF of both instruments is accounted for. As such, we degrade the JWST data to approximate MUSE's spatial resolution by convolving the \paa velocity field from Fig.~\ref{fig:gas_lines} \textbf{b} with a two-dimensional Gaussian kernel with a FWHM of $0.90''$, matching the MUSE seeing, and subsequently reproject the data to the MUSE pixel size ($0.2''$). The resulting map is shown in Fig.~\ref{fig:jwst_binning} and closely resembles the MUSE \ha velocity field: the central blue- and redshifted components become blended, and the distinct CND is no longer resolved. This demonstrates that the apparent simplicity of the MUSE kinematics arises primarily from limited spatial resolution, rather than intrinsic differences in the gas dynamics.

\begin{figure}
    \centering
    \includegraphics[width=\linewidth]{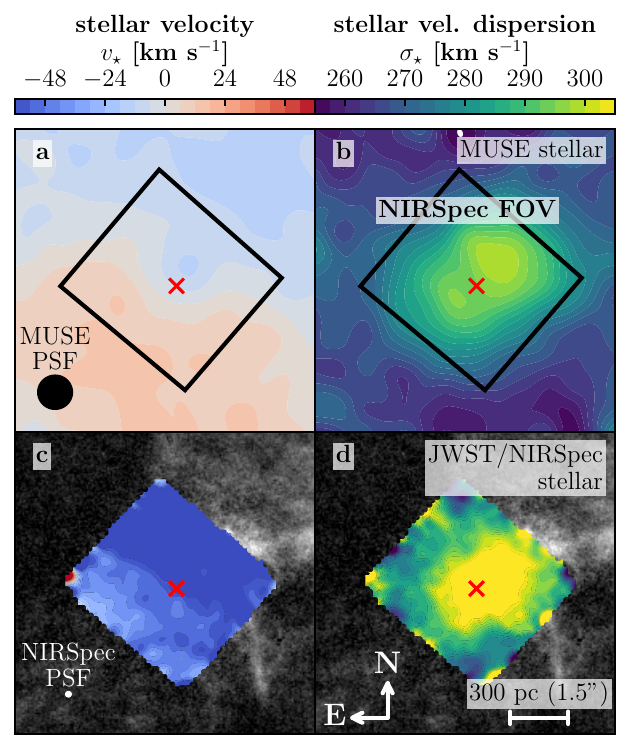}
    \caption{Stellar kinematics obtained with the MUSE observations compared to the JWST/NIRSpec stellar template fitting with LOKI. The colormaps are shared within columns. In the top row, the NIRSpec FOV is shown with the black square on top of the MUSE stellar kinematics, which were convolved with a gaussian kernel having a standard deviation of 1.5 pixels and plotted using contours. In the bottom row, the HST F665N \ha image subtracted by the F814W filter is used as a background for the JWST/NIRSpec stellar maps, which were smoothed using the methods detailed in Section~\ref{sec:Gas Kinematics}. The red cross marks the position of the kinematic centre (see Section~\ref{sec:Black Hole Position}). The PSF FWHM of both instruments is also plotted, with MUSE being seeing-limited ($\sim0.90''$) and NIRSpec having a PSF FWHM of $\sim0.17''$ at \paa's wavelength as an approximation \citep{jones_blackthunder_2026}.}
    \label{fig:muse_stellar_kinematics}
\end{figure}

The stellar kinematics derived from the MUSE data using \texttt{pPXF} \citep{cappellari_parametric_2004} are shown in the top row of Fig.~\ref{fig:muse_stellar_kinematics}, alongside the NIRSpec results (bottom row). Both datasets reveal a weak large-scale velocity gradient, with values from MUSE decreasing from $\sim10$ \kms in the southeast to $\sim-15$ \kms in the northwest. The stellar velocity dispersion maps also show qualitatively similar morphologies, characterized by a smooth and extended peak centred on the kinematic centre.
% However, the velocity dispersion is systematically lower in the MUSE data, reaching values of $\sim280$ \kms compared to $\sim350$ \kms with NIRSpec. This discrepancy is discussed in Section~\ref{sec:Comparison to MUSE}.

\begin{figure*}
    \centering
    \includegraphics[width=\linewidth]{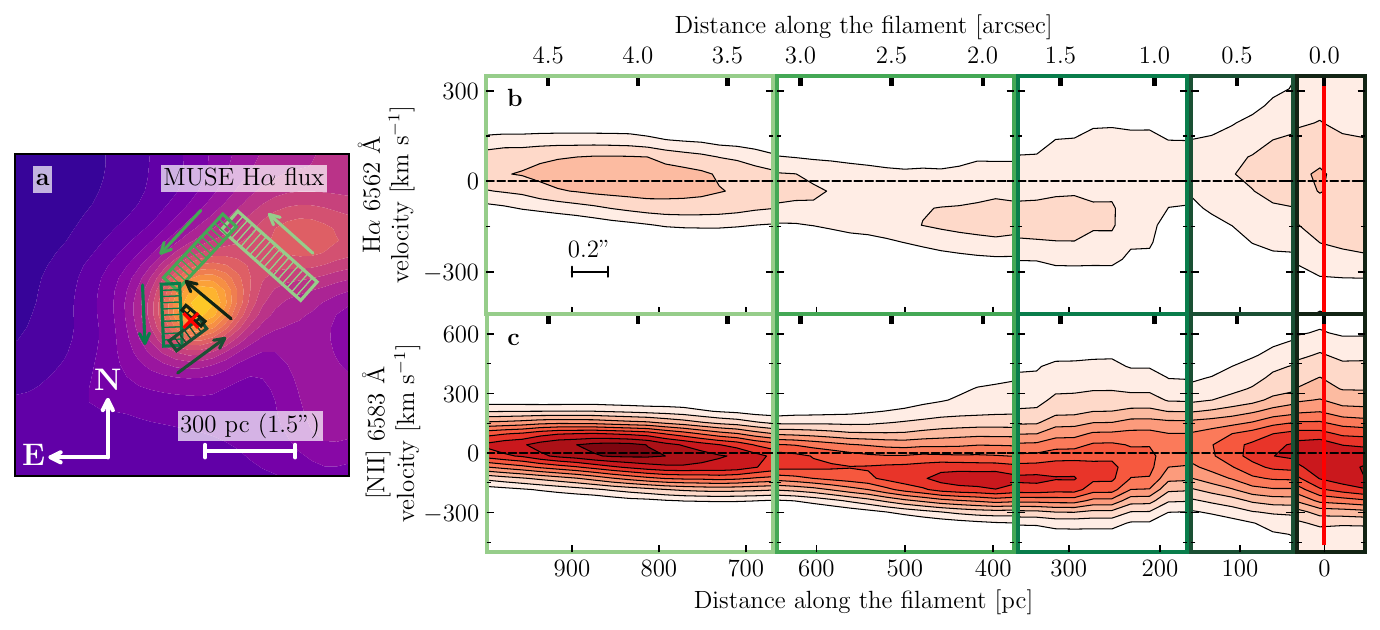}
    \caption{PV diagram of the gas flow along the filament with the MUSE data, showing a smaller velocity gradient and broadening than the NIRSpec data (Fig.~\ref{fig:PV_inflow}). \textbf{a} The \ha flux map on which the PV apertures have been plotted. The first aperture is the lightest one located in \textit{Filament C} at the northwest and the flow goes counter-clockwise following the arrows' directions. The background is the MUSE \ha flux map processed using the methods detailed in Section~\ref{sec:Gas Kinematics}. The red cross marks the position of the kinematic centre (see Section~\ref{sec:Black Hole Position}). \textbf{b} and \textbf{c} feature PV diagrams respectively for the \ha 6562 \AA\ and [\ion{N}{2}] 6583 \AA\ emission lines, obtained by fitting the continuum near these lines with a simple linear function. Contours start at $15\sigma$ and increase in $8\sigma$ increments. The 0 position, highlighted by a vertical red line, corresponds to the position of the kinematic centre, which is in the last (black) aperture. The pixel size of $0.2''$ is also shown in panel \textbf{b}.}
    \label{fig:muse_PV_inflow}
\end{figure*}

To further compare the gas kinematics, we constructed PV diagrams from the MUSE data using the same apertures as those applied to the NIRSpec observations (Section~\ref{sec:Gas Kinematics}). The continuum was modelled and subtracted locally using a linear fit in a line-free range near the \ha emission line, and the noise level ($\sigma$) was estimated from nearby continuum windows to define detection thresholds. The resulting PV diagrams for the \ha and [\ion{N}{2}] 6583 \AA\ emission lines are shown in Fig.~\ref{fig:muse_PV_inflow}. These two lines are successfully separated since the velocity resolution of $\sim120$ \kms is much smaller than the velocity difference between the two lines ($>900$ \kms). In comparison to the NIRSpec results (Fig.~\ref{fig:PV_inflow}) which revealed a \paa velocity gradient of $\sim800$ \kms, the MUSE PV diagrams reveal only faint velocity gradients in the three inner apertures, from $\sim-150$ \kms to $\sim0$ \kms at the kinematic centre's position.

\begin{figure}
    \centering
    \includegraphics[width=0.994\linewidth]{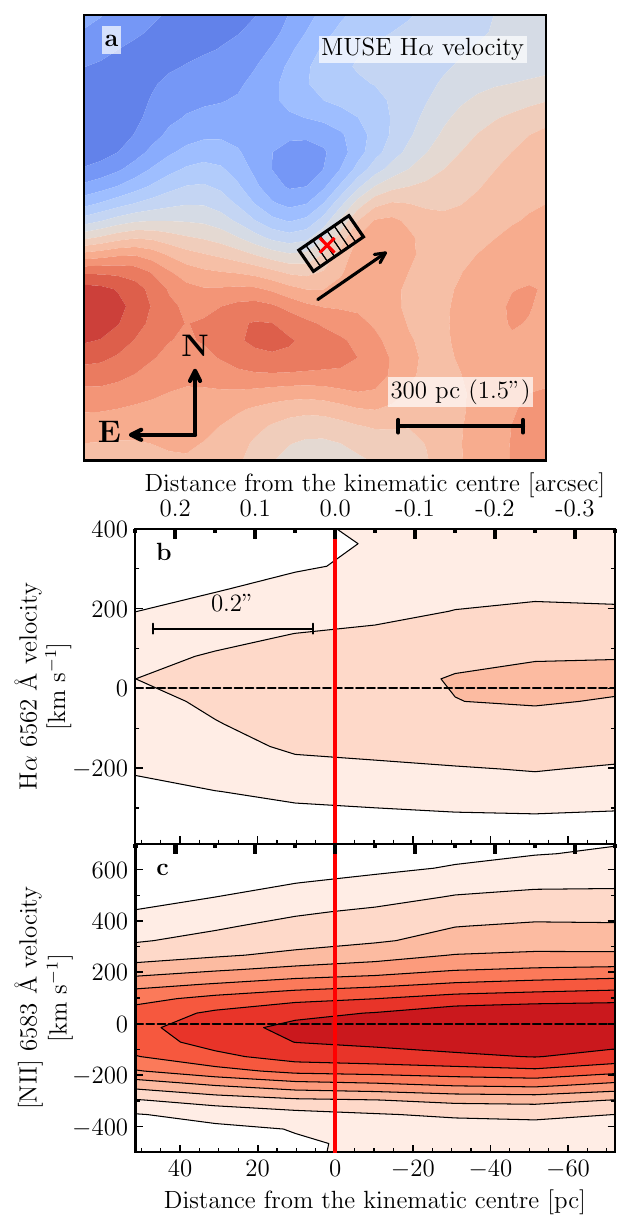}
    \caption{PV diagram of the gas in the plane of the CND, highlighting the absence of a blue-/redshifted structure compared to the NIRSpec data with the same aperture (Fig.~\ref{fig:PV_AGN}). \textbf{a} The \ha velocity map on which the PV aperture has been plotted. The arrow indicates the direction of decreasing distance from the kinematic centre. The background is the MUSE \ha velocity map processed using the methods detailed in Section~\ref{sec:Gas Kinematics}. The red cross marks the position of the kinematic centre (see Section~\ref{sec:Black Hole Position}). \textbf{b} and \textbf{c} feature PV diagrams respectively for the \ha 6562 \AA\ and [\ion{N}{2}] 6583 \AA\ emission lines, obtained by fitting the continuum near these lines with a simple linear function. Contours start at $15\sigma$ and increase in $8\sigma$ increments. The 0 position, highlighted by a vertical red line, corresponds to the position of the kinematic centre and positions after this are negative by convention. The pixel size of $0.2''$ is also shown in panel \textbf{b}.}
    \label{fig:muse_PV_AGN}
\end{figure}

Similarly, PV diagrams extracted along an aperture aligned with the kinematic position angle of the disk (Fig.~\ref{fig:muse_PV_AGN}) show no evidence of a CND or multi-component structure in either \ha or [\ion{N}{2}]. Only a systematic increase in flux and line width can be observed.

\begin{figure*}
    \centering
    \includegraphics[width=\linewidth]{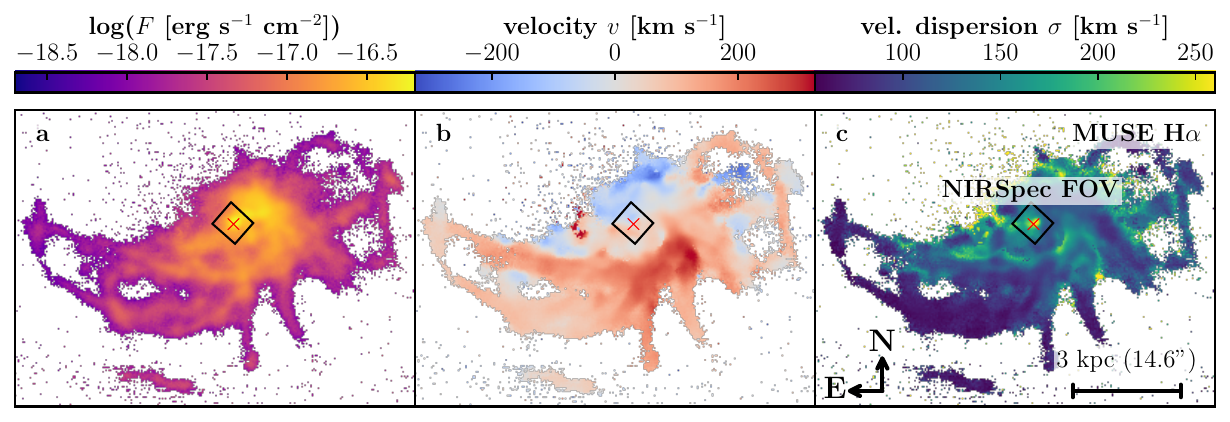}
    \caption{\ha observations by MUSE of the entire NGC~4696 BCG. The logarithmic flux (panel \textbf{a}), velocity (panel \textbf{b}) and velocity dispersion (panel \textbf{c}) are shown. The data has not been smoothed but a threshold of SNR $>5$ was applied. The NIRSpec FOV is shown with the black square and the red cross marks the position of the kinematic centre (see Section~\ref{sec:Black Hole Position}).}
    \label{fig:muse_outer_kinematics}
\end{figure*}

Finally, we exploit the larger $1'\times1'$ FOV of MUSE (much more extensive than NIRSpec's $\sim3''\times3''$) to examine the global gas kinematics of NGC~4696 (Fig.~\ref{fig:muse_outer_kinematics}). The \ha maps reveal an extended network of bright filaments surrounding the central region, with complex velocity structures on kiloparsec scales. The circumnuclear region, corresponding to the region probed by NIRSpec, lies at the location of peak flux and velocity dispersion. On larger scales, the velocity field (panel \textbf{b}) suggests a coherent rotation pattern, with the northern regions predominantly blueshifted and the southern regions redshifted. Additionally, the maps indicate the presence of overlapping kinematic components (e.g. to the east and north-west of the nucleus), possibly corresponding to foreground and background filaments along the line of sight. These large-scale structures provide important context for interpreting the small-scale kinematics revealed by the NIRSpec data, highlighting the multi-phase and multi-scale nature of gas dynamics in the core of NGC~4696, and we discuss these in Section~\ref{sec:Comparison to the Outer Kinematics}.

\subsection{ALMA Observations}\label{sec:ALMA Observations}

\begin{figure}
    \centering
    \includegraphics[width=\linewidth]{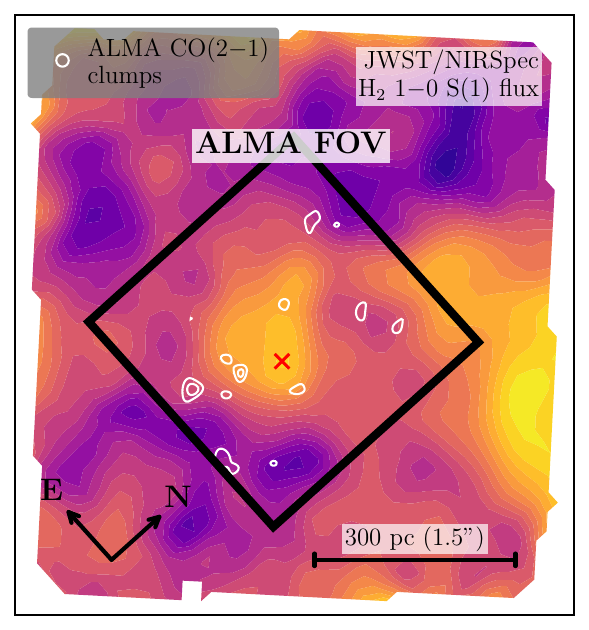}
    \caption{ALMA CO(2--1) detections in NGC~4696, highlighting no extended structure cospatial with the \hs{1} flux. The background is the JWST/NIRSpec \hs{1} flux map fitted using LOKI and processed using the methods detailed in Section~\ref{sec:Gas Kinematics}. The red cross marks the position of the kinematic centre (see Section~\ref{sec:Black Hole Position}). The detected clumps are plotted with white contours drawn at $2\sigma$ and $3\sigma$. The ALMA FOV is also drawn.}
    \label{fig:ALMA}
\end{figure}

\cite{olivares_ubiquitous_2019} present CO(1--0) ALMA observations of NGC~4696, which probe the entire molecular gas structure over $\sim5.6$ kpc, and detect continuum emission with a total flux density of 40.9 mJy. These however have limited spatial resolution (angular resolution of $0.6\times0.4$ kpc$^2$). We therefore compare our NIRSpec results with new ALMA observations of CO(2--1) emission in the core of NGC~4696 ($\sim2''\times2''$, $\sim410\times410$ pc$^2$), shown in Fig.~\ref{fig:ALMA} (Canning et al. in preparation). The detections appear confined to a small number of compact clumps that are barely detected above $2\sigma$. These could be noise peaks or residual sidelobes from the dirty beam in the map due to imperfect calibration, or actual CO(2--1) clumps displaying no clear spatial correspondence to the extended filamentary structures traced by the near-infrared \hs{1} emission or the ionized gas traced by \paa.
% These detections seem to suggest that the molecular gas does not follow the prominent filaments connecting the north-western stellar cluster to the kinematic centre, nor that it exhibits a continuous distribution toward the nucleus.
This mismatch between the potential blobs and molecular/ionized gas emission is discussed in Section~\ref{sec:Comparison to ALMA}.

\section{Discussion}\label{sec:Discussion}
\subsection{Overview}

This study provides a detailed view of the gas and stellar kinematics on sub-kiloparsec scales in the core of a massive BCG hosting strong AGN feedback. The JWST/NIRSpec data reveal coherent and ordered gas motions along the filamentary structures, extending down to a CND. This disk was first reported by \cite{hlavacek-larrondo_jwst_2026} and is studied more in depth here. We also focus exclusively on the kinematics of the gaseous and stellar components, and a complete study of the emission lines is performed by O. Pereira et al. (submitted). The brightest detected lines are \paa, \hs{1}, S(3) and S(5), while no metal lines are detected. Most emission lines are well described by a single kinematic component across the field, with the exception of \paa, which exhibits double-peaked profiles mostly in the inner $\sim100$ pc.

In contrast, the stellar kinematics display a markedly different behaviour. The stellar velocity field is smooth and shows no strong gradient, while the velocity dispersion exhibits a broad, centrally peaked profile. The gas, on the other hand, shows more structured kinematics, including localized enhancements in velocity dispersion near the AGN. This contrast indicates that the gaseous and stellar components are dynamically decoupled and respond differently to the gravitational potential and feedback processes in the central region.

Comparison with complementary datasets further highlights the importance of spatial resolution and tracer selection. The MUSE observations do not recover the small-scale kinematic structures observed in the JWST data, in particular the CND, likely due to seeing-limited resolution. In addition, ALMA observations do not reveal any spatial correspondence between potential CO(2--1) clumps and warmer gas phases, suggesting either a spatial segregation or instrumental limitations.

In the following sections, we examine these results in more detail and discuss their implications for gas accretion, multi-phase structure, and AGN feedback in NGC~4696.

\subsection{Black Hole Properties}
\subsubsection{Black Hole Position}\label{sec:Black Hole Position}

\begin{figure*}
    \centering
    \includegraphics[width=\linewidth]{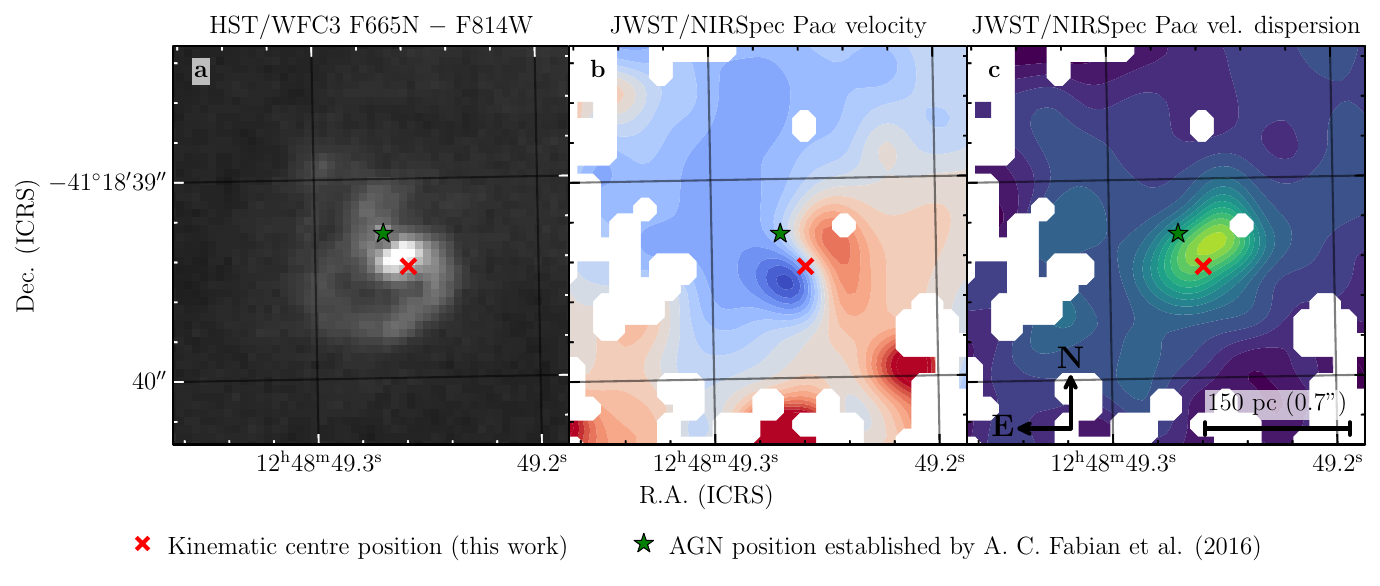}
    \caption{AGN and kinematic centre in NGC~4696, which appear offset by $\sim41$ pc. The green star is the AGN position determined by \cite{fabian_hst_2016} and the red cross is the kinematic centre modelled using our data at ICRS coordinates: $\mathrm{R.A.}=12^\mathrm{h}48^\mathrm{m}49^\mathrm{s}.2583$, $\mathrm{Dec.}=-41^\circ 18'39''.441$. Our estimate agrees with the centre of the CND. \textbf{a} The HST F665N \ha image subtracted by the F814W filter. \textbf{b} and \textbf{c} The JWST/NIRSpec velocity and velocity dispersion fields respectively, processed using the methods detailed in Section~\ref{sec:Gas Kinematics} and filtered for SNRs $>2$.}
    \label{fig:agn_position}
\end{figure*}

\cite{taylor_lowpower_2006} performed 5.0 GHz VLBA observations of the radio source PKS 1246--410 at the centre of NGC~4696 (beam size of $3.74\times1.31$ mas$^2$) and identified a compact source at $\mathrm{R.A.}=12^\mathrm{h}48^\mathrm{m}49^\mathrm{s}.2609$, $\mathrm{Dec.}=-41^\circ 18'39''.417$. They also detected an extended source $\sim25$ pc toward the southwest which they attributed to a one-sided radio jet. In \cite{fabian_hst_2016}, the authors then aligned the AGN and the knot seen with VLBA to similar features seen in the HST images \citep[see Fig.~\ref{fig:agn_position}, but also Fig.~9 of][]{fabian_hst_2016}. However, when the position of the AGN is overlaid on our JWST kinematic maps (see Fig. \ref{fig:agn_position}), it appears offset from the peak in velocity dispersion of the \paa and \hs{1} emission lines, as well as from the centre of the rotating structure. This motivates a reassessment of the AGN position based on the high-resolution kinematic and morphological tracers available in our JWST data.

As mentioned in Section~\ref{sec:Data Collection and Data Reduction}, the astrometric calibration of our NIRSpec data was refined by aligning the swirl structure observed in our data with that seen in archival HST F665N \ha imaging and using the WCS-corrected HST frame with DrizzlePac \citep{anand_drizzlepac_2025} as a reference.

More specifically, in order to estimate the kinematic centre from the \paa velocity field, we follow the procedure described in \cite{GarciaLorenzo_2015}. For a perfect rotational disk-like galaxy, its kinematic centre has a zero rotational velocity, and it is also the location of its largest velocity gradient. We first compute the velocity gradient magnitude map (average directional derivative) at each valid spaxel. After normalizing the gradient to its peak value, we identify the spaxel with the peak gradient as a first estimate of the kinematic centre. Within a region around that peak, we select spaxels whose gradient equals or exceeds the local mean gradient. We choose the reported kinematic centre as the gradient-weighted centroid of the selected spaxels. The established position of the kinematic centre is shown in Fig.~\ref{fig:agn_position} with a red cross and is offset by $\sim41$ pc from the position established by \cite{fabian_hst_2016}, shown with a green star. The derived international celestial reference system (ICRS) coordinates are $\mathrm{R.A.}=12^\mathrm{h}48^\mathrm{m}49^\mathrm{s}.2583$, $\mathrm{Dec.}=-41^\circ 18'39''.441$.

This position is further supported by its consistency with several independent observables. It lies at the centre of the rotating structure identified in other velocity fields, such as \hs{1}, and coincides with the location where the flowing gas appears to reach its highest velocities of $\sim1000$ \kms (as seen in Fig.~\ref{fig:Pa_alpha_gif}). Taken together, these indicators suggest that this position more accurately traces the dynamical centre of the system.

\subsubsection{Black Hole Mass}\label{sec:Black Hole Mass}

The velocity fields shown in Fig.~\ref{fig:gas_lines} reveal organized motions, especially the \paa velocity field in panel \textbf{b} which is consistent with a rotating CND (see Section~\ref{sec:Structures}). Under the simplifying assumption that the gas kinematics are governed by gravitational motions and are close to virial equilibrium, the velocity dispersion can be used to obtain an order-of-magnitude estimate of the SMBH mass.

The maximum \paa velocity dispersion measured in the vicinity of the kinematic centre is $\sigma_{\paa}=449\pm12$ \kms (corresponding to an FWHM of $\sim1057$ \kms). Adopting a characteristic radius corresponding to the pixel size ($0.1''$, or $R\sim20.6$ pc), we estimate the SMBH mass using
\begin{equation}
    M_{\mathrm{BH}}\sim\frac{\sigma_{\mathrm{gas}}^2R}{G}
\end{equation}
where $\sigma_{\mathrm{gas}}$ is our \paa velocity dispersion peak and $G$ is the gravitational constant. This yields an order of magnitude estimate of $M_{\mathrm{BH}}\sim10^9$ \msun.

Previous estimates span a similar range. Using the $M_{\mathrm{BH}}$--$\sigma$ relation and a stellar velocity dispersion of $\sigma\sim254$ \kms, \cite{fabian_hst_2016} also derived $M_{\mathrm{BH}}\sim10^9$ \msun. In contrast, \cite{russell_radiative_2013} used $K$-band luminosity scaling relations based on the 2MASS catalogue \citep{skrutskie_two_2006} and obtained a lower value of $(3.4\pm0.6) \times 10^8$ \msun using relations between the SMBH mass and galaxy properties such as redshift and evolution. This discrepancy may reflect the limited depth of 2MASS imaging which does not allow to encompass accurately the total luminosity of BCGs \citep{lauer_masses_2007}. Similarly, \cite{taylor_lowpower_2006} derived $M_{\mathrm{BH}}\sim3\times10^8$ \msun using the $M_{\mathrm{BH}}$--$\sigma$ relation from \cite{pinkney_kinematics_2003} and a velocity dispersion of $\sigma=262$ \kms \citep{bernardi_redshiftdistance_2002}. \cite{mezcua_most_2018} additionally measured the nuclear X-ray flux from archival Chandra observations and used the relation between SMBH mass and $K$-band 2MASS magnitudes $M_\mathrm{BH}$--$M_\mathrm{K}$ from \cite{graham_mbh_2013} to derive a mass of $M_\mathrm{BH}=10^{9.49\pm0.02}$ \msun. Given the high likelihood of inflowing/outflowing gas in the nucleus, the \paa line width is very likely to be broadened and thus may give an oversized mass. Therefore, we stress that our mass measurement is only an order-of-magnitude estimate, and thus appears to be broadly consistent with the other estimates.

Using our fiducial mass estimate, we can also approximate the SOI of the SMBH as
\begin{equation}
    r_\mathrm{inf}=\frac{GM_{\mathrm{BH}}}{\sigma_\star^2},
\end{equation}
where $\sigma_\star=262$ \kms \citep{bernardi_redshiftdistance_2002} is the stellar velocity dispersion. This yields $r_\mathrm{inf}\sim60$ pc, indicating that the SOI is marginally resolved at our pixel size of $\sim20$ pc, spanning a diameter of $\sim6$ pixels.

A more robust SMBH mass measurement will require detailed dynamical modelling of the rotating structure, which will be presented in a forthcoming study (M. Marquis et al. in preparation).

\subsubsection{Accretion Rate}\label{sec:Accretion Rate}

In Section~\ref{sec:Structures}, we show that the gas appears to be accelerating towards the nucleus. We can thus derive an order-of-magnitude estimate of the gas inflow rate lower bound feeding the CND from the PV diagrams shown in Fig.~\ref{fig:PV_inflow}, which trace the proposed accretion channel linking the filamentary structures, the swirl, and the circumnuclear region. Approximate inflow velocities were estimated by averaging the velocity peaks in each spatial bin of the \hs{1} emission line within the three innermost apertures. This yields characteristic velocities of $v_3\sim-160$ \kms, $v_4\sim70$ \kms, and $v_5\sim470$ \kms for apertures 3--5, respectively (corresponding to the inner $\sim350$~pc from the kinematic centre in Fig.~\ref{fig:PV_inflow}). For the innermost aperture, the velocity was sampled only up to the kinematic centre.

The warm molecular gas mass inside each aperture can then be estimated as
\begin{equation}
    M_{\rm H_2} = N_{\rm tot}\,\Omega\,D_A^2\,2m_{\rm H},
\end{equation}
where $N_{\rm tot}$ is the total column density of warm H$_2$ gas, $\Omega$ is the solid angle of the aperture, $D_A$ is the angular diameter distance and $m_\mathrm{H}$ is the standard mass of a hydrogen atom. We derive these quantities in detail in O. Pereira et al. (submitted).

The resulting gas masses are $M_3\sim\num{4.5e6}$~\msun, $M_4\sim\num{2.1e6}$~\msun, and $M_5\sim\num{1.3e6}$~\msun for apertures 3--5, respectively. As above, the mass in the innermost aperture was computed only up to the kinematic centre. Assuming radial inflow, the mass flow rate lower limit can be approximated as
\begin{equation}
    \dot{M} = \frac{Mv}{r},
\end{equation}
where $r$ is the characteristic length of the aperture and $v$ is the radial velocity computed previously. This yields approximate inflow rates of minimally $\sim3.8$~\msunpyr, $\sim1.3$~\msunpyr, and $\sim18$~\msunpyr through apertures 3--5, respectively. Under these assumptions, the inflow rate lower bound toward the nucleus appears to reach $\dot{M}\sim18$ \msunpyr in the innermost region. However, the level of accretion needed to power the radio source is several orders of magnitude ($\sim3$) smaller than this flow \citep{fabian_hst_2016}, indicating that most of the mass does not reach the SMBH. If low mass stars are formed (see Section~\ref{sec:Global Kinematics of the Stars}), they could also fall into the SMBH without emitting radiation \citep{fabian_hidden_2023}. Moreover, using XMM-Newton Reflection Grating Spectrometer (RGS) data and a multilayer absorption model, \cite{fabian_hidden_2022} measured a hidden cooling flow of 14.2 \msunpyr. The consistency between this value and our estimated lower bound inflow rate of $\sim18$ \msunpyr therefore suggests the validity of the hidden cooling flow hypothesis in NGC~4696. Further detection and mapping of the [\ion{Ne}{6}] line would be needed to establish with certainty the presence of cooling gas to $10^5$ K and below, and thus clearly confirm the hypothesis above \citep{fabian_hidden_2024}.

Comparable inflow rates on similar scales have been reported in NGC~1275. For example, \cite{oosterloo_closing_2024} used ALMA CO(2--1) observations together with a conversion factor of $\alpha_{\rm CO}=0.8$~\msun~K$^{-1}$~km$^{-1}$~s~pc$^{-2}$ to estimate the molecular gas mass in three filaments accreting onto the CND. These filaments are located inside of a region with a radius of $\sim250$ pc. Using infall velocities of order $\sim200$ \kms (consistent as well with our rough estimates for NGC~4696) derived from simulations \citep{lim_radially_2008}, they obtained an accretion rate of $\dot{M}\sim40$~\msunpyr onto the CND, approximately a factor of two higher than the value inferred here for NGC~4696.

We stress that these are very rough estimates: the warm molecular gas mass is obtained by assuming that the gas is described by a single temperature and that it follows a Boltzmann distribution, whereas we approximate infall velocities with radial velocities, which only provide lower bounds for the accretion rate onto the CND.

\subsection{General Gas Kinematics}
\subsubsection{Structures}\label{sec:Structures}
\textbf{Circumnuclear Disk}

In the centre column of Fig~\ref{fig:gas_lines}, we show the velocity fields of the \hs{1}, S(3) and S(5) emission lines as well as the \paa emission line. These velocity fields reveal a blueshifted/redshifted structure consistent with a smoothly rotating CND, especially through \paa (panel \textbf{b}) which is the brightest of our detected lines. To assess whether this structure is dynamically supported by rotation, we consider the ratio $v_c/\sigma$, which compares the ordered rotational velocity to the local velocity dispersion. This diagnostic is commonly used to distinguish rotation-dominated systems ($v_c/\sigma>1$) from dispersion-dominated ones ($v_c/\sigma<1$) \citep[e.g.,][]{forsterschreiber_sins_2009,epinat_integral_2009,genzel_rapid_2006,gaspari_raining_2017}. To calculate the circular velocity, we need to correct the observed velocity difference for the inclination of the disk following
\begin{equation}
    v_c=\frac{v_\mathrm{max}-v_\mathrm{min}}{2\sin i}
\end{equation}
where $v_\mathrm{max}$ and $v_\mathrm{min}$ are the extremal velocities of the CND and $i$ is the inclination angle between the line of sight and the axis of rotation \citep{epinat_integral_2009}. Assuming that the jet axis is aligned with the disk rotation axis, we adopt an inclination of $i\sim70^\circ$ based on the jet orientation reported by \cite{taylor_lowpower_2006}. From the \paa velocity field, we measure $v_\mathrm{max}\sim607$ \kms and $v_\mathrm{min}\sim-800$ \kms, yielding a circular velocity of $v_c\sim750$ \kms. Using the maximum velocity dispersion obtained at the position of the kinematic centre of $\sigma_{\paa}=449\pm12$ \kms, we obtain $v_c/\sigma\sim1.7>1$, indicating that the system is rotation-dominated. This supports the interpretation of the central structure as a dynamically significant and settled CND.

In \cite{hlavacek-larrondo_jwst_2026}, the authors report a disk with a radius of $\sim120$ pc. However, when we measure the spatial separation between the most blueshifted and redshifted emission in the unsmoothed \paa velocity map, we obtain a radius of $\sim60$ pc (roughly 3 pixels), which is consistent with our estimated radius of influence of $r_\mathrm{inf}\sim60$ pc (also a radius of $\sim3$ pixels, see Section~\ref{sec:Black Hole Mass}). Our obtained value thus represents a lower limit on the size of the disk. We can also estimate the warm molecular gas mass enclosed within the CND ($r\sim120$ pc) using the method described in Section~\ref{sec:Accretion Rate}, obtaining $M_{\rm H_2}\sim\num{1.2e7}$ \msun. Combining this estimate with the inflow rate lower bound of $\dot{M}\sim18$ \msunpyr derived in Section~\ref{sec:Accretion Rate} yields an upper bound on the characteristic disk assembly timescale of $\sim0.68$ Myr. This is consistent, to within an order of magnitude, with the free-fall time from the Bondi radius \citep[$\sim0.2$ Myr;][]{fabian_hst_2016}, suggesting that the inferred inflow rate is physically plausible.

A similar CND-like structure has been reported in NGC 1275, the BCG of the Perseus cluster of galaxies \citep{scharwachter_kinematics_2013}. The authors propose that this disk corresponds to the outer regions of a turbulent, collisionally excited accretion disk. Since the rotating motion appears at dynamical equilibrium \cite[see also][]{hlavacek-larrondo_jwst_2026}, the accretion disk likely had time to align with both the CND and the rotation plane of the SMBH as per the Bardeen-Petterson effect \citep{bardeen_lensethirring_1975}, suggesting that the CND corresponds to the outer region of an inner accretion disk. Subsequent studies of the CND in NGC 1275 were performed by \cite{nagai_alma_2019} and \cite{oosterloo_closing_2024} and are reviewed in \cite{hlavacek-larrondo_jwst_2026}.

Additional support for a CND is provided by the PV diagrams (Figs~\ref{fig:PV_inflow} and \ref{fig:PV_AGN}). As noted by \cite{hlavacek-larrondo_hubble_2025}, CNDs are expected to exhibit smooth velocity gradients with minimal pixel-to-pixel variation. This behavior can be seen in the innermost apertures of Fig.~\ref{fig:PV_inflow}, where the velocity varies smoothly by $\sim800$ \kms as the gas approaches the AGN. Moreover, \cite{oosterloo_closing_2024} reported a characteristic velocity twist across the nucleus in similar PV diagrams of NGC 1275. A similar feature is observed in NGC~4696 (Fig.~\ref{fig:PV_AGN}, panel \textbf{b}), where the velocity structure changes sign across the kinematic centre, producing a symmetric twisting pattern about 0 \kms. The broadened line profiles near the kinematic centre likely reflect increased turbulence or non-circular motions induced by the AGN.

% Taken together, the symmetry of the velocity field, the smooth gradients observed in PV space, and the consistency with analogous systems provide strong evidence that the central kinematics of NGC~4696 are dominated by a rotating CND. From Fig.~\ref{fig:gas_lines} \textbf{b}, we can thus establish that this structure depicts a rotation with the eastern (left) side approaching the observer and the western (right) side receding.

% The CND region also features distinct kinematics. In Fig.~\ref{fig:Pa_alpha_multi_component}, we identify 65 spaxels ($\sim7.5\%$ of the pixels with SNR $>2$) exhibiting significant double-component structures according to the criterion described in Section~\ref{sec:Emission Line Fitting}. These detections are primarily concentrated within the CND, inside a radius of $\sim50$~pc from the kinematic centre. The redshifted component generally exhibits larger velocity dispersions, with typical values of $\sim370$ \kms within the central $\sim50$~pc, compared to $\sim250$ \kms for the blueshifted component.

The CND region also features distinct kinematics, with double-component detections of \paa primarily within a radius of $\sim50$ pc and highlighting a broad redshifted component (see Section~\ref{sec:Gas Kinematics}).
Multiple kinematic components have also been observed in the nucleus of M87. The nuclear region of this galaxy is characterized by a rotating disk of ionized gas with an inclination-corrected rotational velocity of $v_c\sim750$ \kms \citep{ford_narrowband_1994} and whose major axis is approximately perpendicular to the jet \citep{harms_hst_1994}. Using MUSE observations, \cite{osorno_revisiting_2023} further identified overlapping velocity structures associated with this rotating disk ($r\sim66$ pc), as well as with the presence of filamentary gas and an AGN-driven outflow. In their interpretation, a biconical outflow produces blueshifted filamentary gas in front of the nucleus and redshifted gas behind it. They also report higher velocity dispersions ($\gtrsim300$ \kms) in the nuclear region, which they attribute to the interaction between outflowing gas and disk rotation.

Although the geometry and line of sight motions in NGC~4696 do not allow us to unambiguously distinguish inflow from outflow, the presence of localized double-component structures near the nucleus, combined with the enhanced velocity dispersions, may indicate the presence of an AGN-driven outflow component. A similar interplay between rotating gas and disturbed nuclear kinematics may therefore be present in NGC~4696. Additionally, such disturbed nuclear kinematics could also naturally arise from the interaction between inflowing filamentary gas and AGN feedback in the central region.

\textbf{Filament}

In Fig.~\ref{fig:regions} we show that the CND is connected toward the northwest by a filamentary structure characterized by enhanced emission. This feature is prominently detected in multiple tracers, including \paa, \hs{1}, S(3), S(5) (Fig.~\ref{fig:gas_lines}) and \ha (Fig.~\ref{fig:muse_gas_lines}).
%The presence of a coherent structure linking large scales to the CND suggests that gas is being transported between the outer regions and the nucleus, potentially tracing inflow or outflow along a well-defined channel \citep{oosterloo_closing_2024}.
This structure appears to connect kiloparsec scales to the CND \citep{hlavacek-larrondo_jwst_2026}.

The filament traced along the apertures in Figs~\ref{fig:PV_inflow} and \ref{fig:Pa_alpha_gif} exhibits kinematic properties consistent with gravitational free fall \citep{lim_radially_2008}. Specifically, the velocity field (Fig.~\ref{fig:gas_lines}) displays a smooth gradient, with velocities increasing toward the nucleus and reaching values of up to $\sim600$ \kms. This behaviour is corroborated by the PV diagram shown in Fig.~\ref{fig:PV_inflow}, which reveals a monotonic increase in velocity ($\sim800$ \kms) with decreasing radius along a scale of $\sim150$ pc. Such trends are consistent with expectations for gas accelerating within the gravitational potential of the host galaxy \citep{wilman_integral_2009}.

At larger radii along this filament ($\gtrsim800$ pc in Fig.~\ref{fig:PV_inflow}, corresponding to the \textit{Filament C} region), the gas exhibits velocities close to systemic, consistent with a scenario in which cooling gas initially resides at rest with respect to the stellar component before being accelerated inward \citep{lim_radially_2008}. The transition from near-systemic velocities at radii of $\gtrsim200$ pc to high velocities of $\sim600$ \kms toward the nucleus strongly favours an infall interpretation, in which gas streams along the filament and feeds the central CND, as expected from chaotic cold accretion \citep{gaspari_raining_2017}. Note however that there may be projection effects which could imply that we are looking at superposed features at different positions along the line of sight.
The observed connection between large-scale filaments and the CND is similar to that reported in the Perseus cluster by \cite{oosterloo_closing_2024}, and this is detailed in \cite{hlavacek-larrondo_jwst_2026}. Additionally, another prediction of chaotic cold accretion is that the kinematics of the ICM should match the kinematics of the ionized and molecular gas, and this is discussed further in Section~\ref{sec:Comparison to XRISM}.

\textbf{Swirl}

Using narrow-band HST imaging of the \ha emission in NGC~4696, \cite{fabian_hst_2016} identified a compact, S-shaped swirl co-spatial with the emission peak near the nucleus, with a characteristic diameter of $\sim1''$. They interpreted this structure as evidence for a connected inflow rather than a chaotic precipitation of cold gas blobs, suggesting that gas near the Bondi radius retains a non-negligible angular momentum and undergoes slow rotation.

In Fig.~\ref{fig:Pa_alpha_gif}, we present continuum-subtracted \paa emission as a function of velocity, revealing a morphology closely resembling the \ha swirl reported by \cite{fabian_hst_2016}. The emission appears to trace a continuous structure linking the large-scale filament to the nucleus through this swirling feature. The PV diagrams shown in Fig.~\ref{fig:PV_inflow}, constructed along apertures that follow the filament and swirl down to the kinematic centre, further support this connection. In particular, the region corresponding to the swirl (primarily encompassed by the fourth aperture) exhibits a smooth and monotonic variation in velocity, with kinematics that aren't chaotic but that connect continuously to both the upstream filament (third aperture) and the inner disk (last aperture) (see also Section~\ref{sec:Witnessing the Infall of Gas onto the SMBH}). This behaviour is observed consistently in both \paa and \hs{1}, indicating that the structure is coherent across multiple gas phases. Together, these results support the interpretation that the swirl represents a dynamical link between the kiloparsec-scale filamentary network and the circumnuclear environment \citep[for a complete review, see][]{hlavacek-larrondo_jwst_2026}.

A similar swirling morphology has been reported in NGC 1275 by \cite{scharwachter_kinematics_2013} in \hs{1} emission, although its origin--whether inflowing or outflowing--remained ambiguous. In contrast, the PV diagrams presented here show evidence of acceleration toward the nucleus, favouring an inflow scenario in NGC~4696. Notably, the velocity evolution along the swirling structure transitions from negative to positive values (Fig.~\ref{fig:PV_inflow}), suggesting that the gas joins the CND on a prograde orbit. Such a configuration is expected to minimize shocks and turbulence, promoting a relatively quiescent accretion process and facilitating the settling of gas into the disk. This contrasts with the interpretation of \cite{scharwachter_kinematics_2013}, who found evidence for retrograde accretion in NGC 1275, a configuration likely to induce stronger shocks and enhance turbulent dissipation, thereby promoting more rapid accretion onto the nucleus.

These observations highlight the importance of angular momentum in regulating accretion onto the SMBH. Rather than purely radial infall, the kinematics indicate that the gas follows more complex trajectories. In NGC 1275, \cite{oosterloo_closing_2024} similarly report disordered, non-axisymmetric motions, with filaments exhibiting significant angular momentum and following non-radial paths toward the centre. Such behaviours are consistent with chaotic cold accretion models, in which gas condenses out of the hot phase and accretes along filamentary streams with non-zero angular momentum, often tracing helical or conical trajectories \citep[e.g.][]{gaspari_raining_2017}. The swirling structure observed in NGC~4696 appears to be a direct manifestation of this process.

\textbf{Alignment with the jet}

In \cite{hlavacek-larrondo_jwst_2026}, the authors report that the sub-kiloparsec scale jet appears broadly perpendicular to the CND.
% In Fig.~\ref{fig:gas_lines} \textbf{b}, the \paa velocity field reveals a CND, as discussed in Section~\ref{sec:Structures} Circumnuclear Disk. We determine the orientation of the CND from the axis connecting the peak blueshifted and redshifted emission, and find that the disk lies approximately along the northwest/southeast direction. Comparing this geometry with the jet orientation determined by \cite{taylor_lowpower_2006}, we find that the jet axis is broadly aligned with the inferred rotation axis of the disk, and therefore approximately perpendicular to its plane.
% A similar geometric configuration has been reported in NGC 1275 by \cite{scharwachter_kinematics_2013}, who found that the radio jet is oriented perpendicular to the rotating molecular disk. They argue that this geometry disfavours a jet-driven origin for the observed circumnuclear kinematics traced by the \hs{1} and [\ion{Fe}{2}] lines, as jet-induced motions would be expected to align with the jet axis rather than with the disk plane. Instead, the kinematics are more naturally explained by rotation within a disk.
This alignment suggests a physical connection between the circumnuclear gas and the central engine. \cite{scharwachter_kinematics_2013} interpret the observed disk in NGC 1275 as the outer extension of a smaller-scale accretion disk responsible for launching the jet. This interpretation is supported by ALMA observations of cold molecular gas \citep{nagai_alma_2019}, which reveal a velocity gradient (i.e. a CND) aligned with the jet axis. However, if jet production is governed by the \cite{blandford_electromagnetic_1977} mechanism, the jet direction may instead trace the spin axis of the SMBH, implying that its spin is aligned with the CND.
% Additional support for this interpretation comes from \cite{oosterloo_closing_2024}, who argue that the planar distribution of gas around the CND is more consistent with infall than with jet-driven outflows.

A similar scenario is likely in NGC~4696. The perpendicularity between the CND and the jet reported by \cite{hlavacek-larrondo_jwst_2026} suggests that the observed filamentary structures are not driven by jet-gas interactions. Instead, the geometry supports a picture in which the CND is dynamically connected to the accretion process, either as the outer extension of an accretion disk or as a structure aligned with the SMBH spin axis, which governs the jet orientation.

Such an orthogonal configuration is not unique to NGC~4696, but has been observed in several BCGs. For example, \cite{hamer_cold_2014} found that the velocity gradient of the cold gas in Hydra A is oriented perpendicular to the axis of the radio-inflated cavities. Similarly, in NGC 4261, \cite{jones_radio_2000} observed that the radio jets are perpendicular to a CND and that they maintain a consistent orientation from parsec to kiloparsec scales, implying that the spin axes of the accretion disk and SMBH remain unchanged for at least $10^6$ yr.

The consistency of this geometry across systems suggests that such alignment may be a common outcome of the accretion process in massive galaxies, reinforcing the interpretation that the circumnuclear gas in NGC~4696 traces a dynamically stable structure linked to the central engine.

\subsubsection{Witnessing the Infall of Gas onto the SMBH}\label{sec:Witnessing the Infall of Gas onto the SMBH}

The animated Fig.~\ref{fig:Pa_alpha_gif} provides a direct visualization of the kinematic evolution of the gas as it enters the CND and approaches the nucleus. The systematic shift in the location of the \paa emission with velocity reveals a coherent progression from the filamentary structures toward the nucleus. This behaviour offers a rare view of gas dynamics on sub-kiloparsec scales in the immediate vicinity of a SMBH, while not directly revealing how the gas is ultimately accreted by the SMBH.

A complementary perspective on this apparent acceleration is provided by the PV diagrams tracing the inflow apertures for \paa and \hs{1} (Fig.~\ref{fig:PV_inflow}). By tracking the velocity corresponding to the peak emission at each position along the final apertures, we can quantify the change in velocity as a function of projected distance from the kinematic centre.

\begin{figure}
    \centering
    \includegraphics[width=\linewidth]{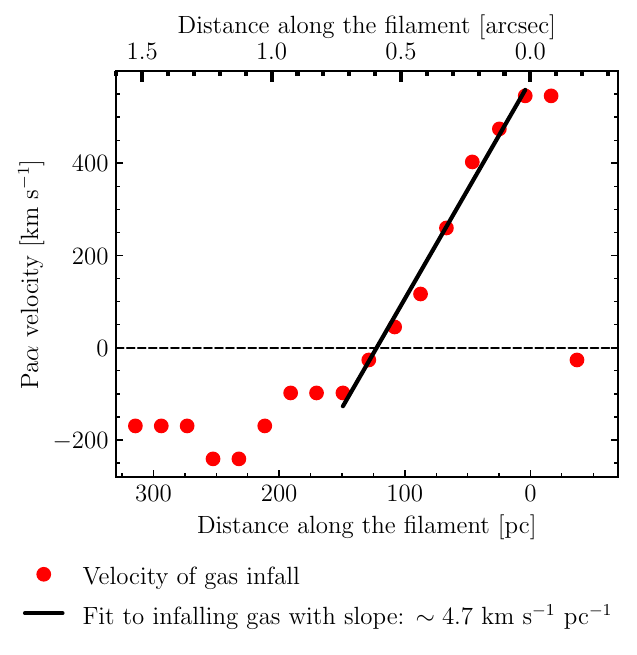}
    \caption{Velocity of the gas as a function of distance along the filament, showing a velocity gradient on scales of $\sim150$ pc. Red points indicate the velocity corresponding to the peak emission in each spatial bin of the PV diagram shown in Fig.~\ref{fig:PV_inflow}, with matching ticks for direct comparison. The inner region ($\lesssim150$ pc) is well described by a linear fit ($R^2=0.986$), corresponding to a velocity gradient of $4.7$ \kms pc$^{-1}$.}
    \label{fig:inflow_velocity_profile}
\end{figure}

Fig.~\ref{fig:inflow_velocity_profile} shows the resulting velocity profile, constructed by identifying the velocity of maximum flux in the PV diagrams over the region where the acceleration is most prominent ($\sim300$ pc from the kinematic centre). A linear fit to the inner region (150 to 0 pc from the kinematic centre) provides an estimate of the velocity gradient, yielding $4.7$ \kms pc$^{-1}$ ($R^2=0.986$).

This measurement indicates a rapid increase in velocity toward the nucleus, consistent with accelerated inflow. Such a direct characterization of the spatially resolved velocity increase along an inflowing structure at these scales remains rarely observed, highlighting the unique capability of high-spatial resolution IFUs like JWST/NIRSpec to probe gas flow in the immediate environment of nearby SMBHs.

\subsubsection{Comparison to ALMA}\label{sec:Comparison to ALMA}

In Fig.~\ref{fig:ALMA}, we present the ALMA CO(2--1) observations of the central $\sim410\times410$ pc$^2$ of NGC~4696 overlaid on the JWST \hs{1} flux map. No extended CO emission is detected, with only a handful of clumps detected at a 2--3$\sigma$ level which could also be noise peaks. This may indicate a lack of a cold molecular gas reservoir traced by CO(2--1) in the central region or limitation from the array configuration.

As previously mentioned, \cite{olivares_ubiquitous_2019} reported extended CO(1--0) emission that is broadly co-spatial with the brightest regions of the warm ionized nebula. The absence of corresponding CO(2--1) detections in our data is therefore unexpected and may result from the antenna configuration, whose \textit{uv} coverage is not optimized for recovering diffuse, extended emission. Consequently, our observations do not demonstrate the absence of CO(2--1) in the nuclear region, but only that no significant emission is detected in the present dataset.

Whether this non-detection is observational or reflects a genuine deficit of cold molecular gas will be investigated in a future study using new ALMA observations targeting molecular clumps at a resolution comparable to JWST's. Such observations will resolve the cold molecular gas within the Bondi radius while simultaneously recovering the extended molecular component, enabling a direct comparison between the spatial distribution and kinematics of the CO, warm molecular, and ionized gas phases.

If the present non-detection ultimately proves to be physical, this would be particularly noteworthy given that CO is the primary tracer of the cold molecular gas reservoir \citep[e.g.][]{edge_detection_2001}, and is typically found to correlate spatially with warm molecular and ionized gas in BCGs \citep[e.g.][]{salome_cold_2003,hamer_relation_2012}. The presence of strong H$_2$ and \paa emission in NGC~4696 therefore may suggest a discrepancy between different gas tracers.

\subsubsection{Comparison to XRISM}\label{sec:Comparison to XRISM}

Previous deep \textit{Chandra} observations by \cite{sanders_very_2016} revealed strong signatures of gas sloshing, driven by the interaction between two subclusters within the Centaurus cluster \citep[see][]{lucey_centaurus_1986}. More recently, high-resolution X-ray spectroscopy using XRISM/Resolve by \cite{xrismcollaboration_bulk_2025} revealed a bulk velocity of $-128\pm6.3$ \kms and a turbulent velocity dispersion of $117\pm9$ \kms within the inner $\sim10-20$ kpc. Our observations target a much smaller scale, essentially the inner $\sim620\times620$ pc$^2$, and we are therefore unable to directly compare the detailed blueshifted/redshifted structure of the CND to the kinematics of the hot X-ray gas.

Nonetheless, the average velocity dispersion of both \paa components in the \textit{Whole} region (Fig.~\ref{fig:regions}) is $\left<\sigma_{\paa}\right>\sim130$ \kms. The similarity between the velocity dispersion of the molecular gas and that of the hot ICM, despite the vastly different spatial scales probed, namely $\sim3''\times3''$ with NIRSpec versus more than $1'\times1'$ for the ``centre" region indicated in \cite{xrismcollaboration_bulk_2025} Fig.~1 \textbf{b}, may suggest a degree of dynamical coupling between the two phases. In terms of bulk velocity, even if probing vastly different scales, the hot X-ray gas is decoupled from the \paa and molecular H$_2$ gas phases tracing hundred to thousand degree gas. Indeed, the average bulk velocity of the two \paa components in the \textit{Whole} region fit (Fig.~\ref{fig:regions}) is roughly $\sim-44$ \kms compared to $\sim-128\pm6.3$ \kms for the X-ray gas.

% However, averaging the velocity dispersions of the two \paa components in the \textit{Whole} region fit (Fig.~\ref{fig:regions}), we obtain a characteristic dispersion of $\left<\sigma_{\paa}\right>\sim130$ \kms. The similarity between the velocity dispersion of the molecular gas and that of the hot ICM, despite the vastly different spatial scales probed, namely $\sim3''\times3''$ with NIRSpec versus more than $1'\times1'$ for the ``centre" region indicated in \cite{xrismcollaboration_bulk_2025} Fig.~1 \textbf{b}, may suggest a degree of dynamical coupling between the two phases. In terms of bulk velocity, even if probing vastly different scales, the hot X-ray gas is decoupled from the \paa and molecular H$_2$ gas phases tracing hundred to thousand degree gas. Indeed, averaging the bulk velocity of the two \paa components in the \textit{Whole} region fit (Fig.~\ref{fig:regions}) gives a bulk velocity of roughly $\sim-44$ \kms compared to $\sim-128$ \kms for the X-ray gas.

\subsubsection{Comparison to the Outer Kinematics}\label{sec:Comparison to the Outer Kinematics}

The large-scale structure and assembly history of NGC~4696 have been investigated in several studies. Using \textit{Magellan}/MegaCam photometry, \cite{federle_turbulent_2024} identified 3818 globular cluster candidates and analyzed their spatial distribution. They found that the number and distribution of globular clusters are consistent with a complex merger history (see their Fig.~2), as expected for a BCG residing at the centre of a deep gravitational potential. This supports the view that NGC~4696 has undergone significant interactions and accretion events, which likely influence its present-day gas dynamics. Additionally, using some of the MUSE observations employed in this work, \cite{hamer_discovery_2019} identified several extended structures in the [\ion{N}{2}] 6583 \AA\ emission. They found diffuse [\ion{N}{2}] emission surrounding the optical filaments and filling the regions between them. They also reported a shell-like structure north of the BCG that is spatially coincident with an X-ray shock front, together with a diffuse halo component whose velocity is offset by $\sim100$--300 \kms\ relative to the neighboring filaments.

Although the spatial resolution of the MUSE observations are insufficient to resolve the detailed structure of the nucleus, its wide FOV ($1'\times1'$) enables a characterization of the large-scale ionized gas kinematics. In Fig.~\ref{fig:muse_outer_kinematics}, we present the \ha emission across the full extent of the galaxy with a pixel size of $\sim41$ pc ($0.2''$). The data reveal extended nebular structures on scales of $\sim5$--10 kpc, significantly larger than those probed by the JWST/NIRSpec observations. As a recombination line, \ha traces diffuse ionized gas, which is expected to be more influenced by the AGN than the denser molecular gas that can be traced by JWST \citep{hatch_ionized_2007}.

The corresponding MUSE velocity field (Fig.~\ref{fig:muse_outer_kinematics} \textbf{b}) shows smoothly varying velocities across the filaments and into the central region, suggestive of a coherent large-scale rotation as shown by the red and blue components. In contrast to the steep gradients ($\sim800$ \kms) observed on scales of $\sim150$ pc with NIRSpec (see Fig.~\ref{fig:PV_inflow}), the filaments exhibit much shallower velocity variations ($\sim350$ \kms) on much larger scales ($\sim2$ kpc), consistent with either slow inflow or large-scale rotation. This complex large-scale rotation is not evident in the NIRSpec data, likely because the kinematics of the molecular gas in the immediate vicinity of the SMBH are dominated by AGN-driven processes and ordered circumnuclear flow, rather than by the galaxy’s global dynamics \citep{hlavacek-larrondo_hubble_2025}. However, the alignment of redshifted emission (which peaks at $\sim350$ \kms, $\sim1.9$ kpc toward the southwest) and blueshifted emission (peaking at $\sim-200$ \kms, $\sim1.4$ kpc toward the north) as seen with MUSE is broadly consistent with the orientation of the inner molecular gas infalling onto the CND from our JWST/NIRSpec data, which shows the blueshifted side of the disk extending towards the north and the redshifted side extending towards the south. This suggests a degree of continuity between large- and small-scale flows. Furthermore, the MUSE velocity dispersion map (Fig.~\ref{fig:muse_outer_kinematics} \textbf{c}) shows a gradual increase toward the centre, peaking near the AGN and kinematic centre locations. This trend mirrors the small-scale velocity dispersion behaviour observed in the NIRSpec data, indicating that the kinetic flow around the nucleus may still be linked to the global gas reservoir \citep{hlavacek-larrondo_hubble_2025}.

% On larger scales, X-ray observations provide complementary constraints on the dynamics of the hot phase. Using XRISM/Resolve spectroscopy, \cite{xrismcollaboration_bulk_2025} measured bulk motions in the ICM and found that the bulk flow is not correlated with the motion of ionized or molecular gas. In particular, they report enhanced bulk velocities in regions offset from the core, indicating that the dominant large-scale motions are not driven by the AGN. Instead, these flows are likely associated with sloshing induced by past merger activity, as discussed previously. Similarly, the X-ray cavities were found to spatially correlate with the molecular filaments \citep{olivares_ubiquitous_2019}. The apparent kinematic decoupling between the hot ICM and the cooler phases of ionized and molecular gases suggests that the latter may have condensed from the hot medium and subsequently settled into the central regions, where they evolve under different dynamical conditions \citep{xrismcollaboration_bulk_2025}. In Section~\ref{sec:Global Kinematics of the Stars}, we also show that the stellar kinematics are decoupled from those of the molecular and ionized gas.

\subsection{Global Kinematics of the Stars}\label{sec:Global Kinematics of the Stars}

\begin{figure}
    \centering
    \includegraphics[width=\linewidth]{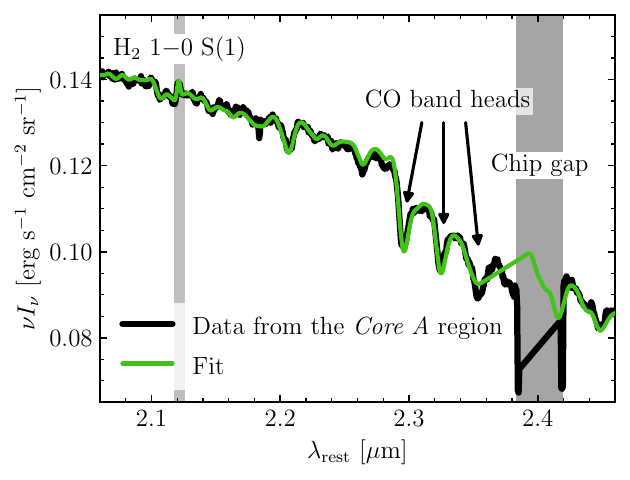}
    \caption{JWST/NIRSpec spectrum of the \textit{Core A} region, featuring distinct CO band heads. The fit using LOKI is shown in green. Three prominent CO band heads from 2.3 to 2.36 \um are identified. The wide dark grey band shows the chip gap in the data, i.e.\ the physical gap between the NIRSpec detectors which prevents collecting data around 2.38--2.42 \um in this region. We also note that a second chip gap can be seen in the fit (green line) around 2.36--2.4 \um, which stems from the fact that the continuum is modelled from the \textit{Background} region, whose chip gap lies in this wavelength range (see Section~\ref{sec:Emission Line Fitting} for a description of the fitting method).}
    \label{fig:co_band_heads}
\end{figure}

In Fig.~\ref{fig:co_band_heads} we show the integrated spectrum from the \textit{Core A} region which highlights pronounced CO band heads in the 2.3--2.36 \um range. This provides a rare opportunity to study in depth the kinematics of the stars, as AGN are known to reduce the depth of CO absorption features \citep{imanishi_nearinfrared_2004}.

In Fig.~\ref{fig:stellar_kinematics}, we show the extracted stellar kinematics from the continuum fit in each spaxel of the JWST observations and compare them to the \paa kinematics. The stellar velocity field (panel \textbf{c}) shows no clear evidence of ordered rotation, in contrast to the coherent structures observed in the gas component especially with \paa (panel \textbf{a}). Instead, it appears largely uniform, with only a weak gradient from the southeast ($\sim -30$ \kms) to the northwest ($\sim -70$ \kms). This variation lies below the instrumental velocity resolution of $\sim110$ \kms, preventing on its own a robust assessment of whether an intrinsic gradient is really present.

However, the MUSE data reveals a weak gradient with the same orientation, from $\sim10$ \kms to $\sim-15$ \kms (see Section~\ref{sec:MUSE Kinematics}). Although it remains below the instrumental spectral resolution ($\sim80$--$170$ \kms), it is unlikely that a gradient of the same orientation would appear due to systematics from two completely different instruments. The fact that this gradient from southeast to northwest is aligned between the JWST/NIRSpec and MUSE data thus suggests it is genuine.

% Moreover, such a small apparent gradient may arise from spectral calibration uncertainties across spaxels, and thus its physical significance remains uncertain\footnote{For more details on NIRSpec's calibration, see the user documentation at \url{https://jwst-docs.stsci.edu/jwst-calibration-status/nirspec-calibration-status/nirspec-calibration-concept}.}.

The NIRSpec stellar velocity dispersion map (Fig.~\ref{fig:stellar_kinematics} \textbf{d}) exhibits a broad profile that peaks near the AGN location at $\sim370$ \kms, lower than the peak dispersion observed in the gas ($\sim450$ \kms), while MUSE (Fig.~\ref{fig:muse_stellar_kinematics} \textbf{b}) shows a central increase reaching $\sim300$ \kms. A direct comparison between the stellar and gas velocity dispersions is shown in Fig.~\ref{fig:sigma_comparison}, where the normalized \hs{1} dispersion is overplotted on the normalized stellar dispersion, along with cuts along the north--south and east--west directions. While both components peak at similar spatial locations, their radial behaviors differ markedly: the gas dispersion declines rapidly over scales of $\sim100$ pc, whereas the stellar dispersion decreases more gradually with radius.

These differences highlight the distinct dynamical states of the stellar and gaseous components. The weakness of ordered stellar motion, together with the relatively smooth dispersion profile, indicates that the stellar motions are dominated by random velocities, as expected for a massive elliptical galaxy \citep{hamer_optical_2016}. In contrast, the gas exhibits coherent kinematic structures, implying that the two components are dynamically decoupled. The broad stellar velocity dispersion is also consistent with the deep gravitational potential of the system, as expected from virial equilibrium.

Such decoupling between stellar and gaseous kinematics has been reported in other BCGs. For example, \cite{olivares_gas_2022} analyzed MUSE observations of 18 BCGs at $z \leq 0.017$ in order to study their kinematics and distribution of optical emission-line gas. They found that the stellar kinematics are often rotation-dominated, while the ionized gas in the filamentary sources exhibits disturbed motions on kiloparsec scales. In most cases, the gas is kinematically decoupled from the stellar component, suggesting an external origin. Similarly, \cite{ciocan_vltmuse_2021} studied the BCG in MACS 1931.8$-$2635 ($z=0.35$) using archival MUSE observations to compare its gas and stellar components. They found that the gaseous kinematics in the core (i.e. on scales of tens of parsecs) are decoupled from the stellar motions, indicating that the gas does not trace the stellar gravitational potential. The observed high stellar velocity dispersion profiles in MACS 1931.8$-$2635 also support a scenario in which the cold gas condenses from the hot CGM rather than originating from the stellar population. In this framework, the similar broad stellar dispersion profile observed in NGC~4696 (Fig.~\ref{fig:sigma_comparison}) is consistent with expectations from precipitation-driven accretion \citep{voit_regulation_2015}, chaotic cold accretion models \citep{gaspari_chaotic_2013} and stimulated feedback \citep{mcnamara_mechanism_2016}, in which the ionized and molecular gas -- and not the stars -- precipitate out of the ICM, meaning that their dynamics are governed by the larger-scale hot atmosphere rather than the stellar component.

% Finally, we explore in Fig.~\ref{fig:PV_absorption_inflow} the stellar component using PV diagrams of two absorption lines extracted along the same filament aperture as in Figs~\ref{fig:PV_inflow} and \ref{fig:muse_PV_inflow}. These track the depth and equivalent width of each absorption feature as a function of position along the apertures. From only the apparent increase in absorption depth toward the nucleus, we could interpret this as reflecting a bottom-heavy initial mass function associated with the cooling flow along the filament. For instance, \cite{fabian_consequences_2024} propose that high thermal pressure environments in cooling flows may favour low-mass star formation, potentially leading to an accumulation of low-mass stars, and thus provoking the observed increase in depth \citep[see also][]{fabian_hidden_2024}. However, since the equivalent width does not simultaneously rise as seen in panel \textbf{d}, this interpretation is uncertain as this could simply be due to a greater surface density of the stars towards the nucleus. Another possible interpretation of this increase in depth is that this occurs from enhanced interstellar medium absorption, caused by increased column density in the nuclear region. To determine the accurate interpretation, the stellar population in NGC~4696 will be explored in a future paper.
Finally, we explore in Fig.~\ref{fig:PV_absorption_inflow} the stellar component using PV diagrams of two absorption lines extracted along the same filament aperture as in Figs~\ref{fig:PV_inflow} and \ref{fig:muse_PV_inflow}. These track the depth and equivalent width of each absorption feature as a function of position along the apertures. \cite{fabian_consequences_2024} propose that high thermal pressure environments in cooling flows may favour low-mass star formation, potentially leading to an accumulation of low-mass stars \citep[see also][]{fabian_hidden_2024}. According to this idea, we should expect an increase in equivalent width, reflecting a bottom-heavy initial mass function associated with the cooling flow along the filament. However, no such increase is seen (see Fig.~\ref{fig:PV_absorption_inflow} \textbf{d}). The fact that the absorption depth increases (panels \textbf{b} and \textbf{c}) is likely caused by a greater surface density of the stars towards the nucleus or by enhanced interstellar medium absorption caused by increased column density in the nuclear region. Further studies would be needed to constrain the stellar population in NGC~4696.

\subsection{Redshift}

The redshift used throughout this paper ($z\sim0.01003$) has been obtained from the stellar component modelled in a collection of MUSE observations \citep{xrismcollaboration_bulk_2025}. They also used Resolve to determine that the ICM emission has heliocentric redshifts of $z\lesssim0.0095$, indicating a bulk blueshift relative to the BCG.
% To obtain the redshift of NGC~4696, \cite{xrismcollaboration_bulk_2025} used a collection of 22 MUSE observations from 2014 to 2023 amounting to a total exposure time of 5.66 hrs and combined the data within the central 1 kpc radius into a single spectrum. This spectrum was then fit with pPXF \citep{cappellari_parametric_2004} to obtain a stellar velocity of $3008\pm7$ \kms, which corresponds to $z\sim0.01003$.
The stellar velocity fields obtained with both JWST and MUSE (see Fig.~\ref{fig:muse_stellar_kinematics}) vary within ranges well below their respective spectral resolutions (see Section~\ref{sec:Global Kinematics of the Stars}), indicating no significant deviation from this systemic value and supporting a stellar component consistent with being at rest in the adopted reference frame \citep{cappellari_parametric_2004}.

% Determining the systemic velocity of the gas is more challenging due to its complex and multi-component kinematics. Indeed, the \hs{1}, S(3), S(5) and especially \paa emission lines displayed in Fig.~\ref{fig:gas_lines} reveal a rotating CND, and Fig.~\ref{fig:Pa_alpha_multi_component} displays the multi-component nature of the \paa emission. However, regions along the filament that exhibit velocities close to zero and which are likely less affected by nuclear inflow appear broadly consistent with the adopted stellar redshift. For example, the \textit{Filament C} displays a velocity offset of $\sim-10$ \kms for the \paa line (see Fig.~\ref{fig:gas_lines} \textbf{b}). This is also apparent in Fig.~\ref{fig:PV_inflow}, which shows that the \paa and \hs{1} velocity distributions overlap with the 0 \kms line.

Assuming that systematic uncertainties are negligible, we can derive an independent estimate of the stellar redshift from our data that would represent the systematic redshift of the stars within the central $\sim620\times620$ pc$^2$. Excluding edge pixels affected by potential artifacts, the average heliocentric/uncorrected velocity is $v_\star = 2949\pm15$ \kms (statistical uncertainty corresponding to the standard deviation of the fitted stellar velocities distribution), coresponding to a redshift of $z_\star = 0.00984\pm0.00005$, which is slightly lower than the the $z=0.01003$ reported by \cite{xrismcollaboration_bulk_2025}. Similarly, for the gas in the \textit{Filament C} region, in which we believe that the gas is not yet falling in (see Section~\ref{sec:Structures} for a discussion on the PV diagrams), we calculate a heliocentric/uncorrected velocity of $v_{\paa}=2995\pm5$ \kms and a redshift of $z_{\paa}=0.00999\pm0.00002$.
All redshifts obtained and discussed are summarized in Table~\ref{tab:redshifts}.

\begin{table*}
    \caption{Derived redshifts for NGC~4696.}
    \centering
    \begin{tabular}{cccr}
        \hline
        Redshift & System & Telescope/instrument & Reference \\
        \hline
        $0.00999\pm0.00002$ & \paa emission in \textit{Filament C} & JWST/NIRSpec & This work \\
        $0.00984\pm0.00005$ & Stellar component & JWST/NIRSpec & This work \\
        $0.01003$ & Stellar component & MUSE & \cite{xrismcollaboration_bulk_2025} \\
        $\lesssim0.0095$ & ICM & XRISM/Resolve & \cite{xrismcollaboration_bulk_2025} \\
        % $0.00987\pm0.00005$ & PKS 1246--410 radio source & VLBA & \cite{taylor_lowpower_2006} \\
        $0.0104$ & Cen 30 subcluster & UK Schmidt-Telescope & \cite{lucey_centaurus_1986} \\
        \hline
    \end{tabular}
    \label{tab:redshifts}
\end{table*}

\section{Conclusions}\label{sec:Conclusions}

We present JWST/NIRSpec IFU observations of the central $\sim620\times620$ pc$^2$ of NGC~4696, the BCG of the Centaurus cluster. These data enable spatially resolved mapping of the kinematics of both ionized and molecular hydrogen through multiple emission lines, with the brightest detections in \paa, \hs{1}, S(3), and S(5). The pixel size ($\sim20$ pc) allows us to probe gas dynamics on scales comparable to the sphere of influence (SOI) of the SMBH, providing a detailed view of gas accretion processes in the nuclear region. The data reveal a network of filamentary structures connecting to a compact circumnuclear disk (CND), along with a swirling morphology of ionized gas in the immediate vicinity of the nucleus. Our main results are summarized as follows:

(i) Consistent with \cite{hlavacek-larrondo_jwst_2026}, we identify a well-defined rotating gaseous structure, referred as a CND, which we characterize using simple dynamical models. We define a kinematic centre associated with this disk, which is spatially offset by $\sim41$ pc from the radio core identified by \cite{taylor_lowpower_2006}. The structure exhibits a circular velocity of $\sim750$ \kms, is rotation-dominated ($v_c/\sigma \sim 1.7$), and appears to have a mass of roughly $\sim\num{1.2e7}$ \msun.

(ii) A pronounced peak in velocity dispersion is observed at the kinematic centre. Interpreting this feature as arising from the gravitational influence of a SMBH yields a mass of $\sim10^9$ \msun and a corresponding SOI of $r_\mathrm{inf}\sim60$ pc, which is resolved by our observations.

(iii) The filamentary gas is consistent with gravitational free fall, displaying a smooth increase in velocity toward the nucleus by $\sim800$ \kms. The filament appears physically connected to the CND, both morphologically and kinematically. In particular, the ionized gas traced by \paa exhibits a velocity gradient of 4.7 \kms pc$^{-1}$ within the inner 150 pc. We also measure a lower bound on the inflow rate in the CND of $\dot{M}\sim18$~\msunpyr assuming local thermal equilibrium and approximating infall velocity with radial velocity, suggesting a maximum assembly timescale of $\sim0.68$ Myr.

(iv) Using a continuum-subtracted data cube, we directly trace the spatial progression of the inflowing gas (see Fig.~\ref{fig:Pa_alpha_gif}). The \paa emission reveals a coherent shift in the location of peak emission with velocity, providing a rare view of accelerating gas as it approaches the kinematic centre.

(v) Comparison with MUSE observations in the central $\sim1.6\times1.6$ kpc$^2$ ($\sim7.6''\times7.6''$) shows that the rotating structure and kinematic substructure detected with NIRSpec are not recovered at lower spatial resolution. Degrading the NIRSpec data to MUSE-like conditions offers a direct view of this phenomenon.
% The MUSE velocity dispersion maps are broader and smoother. Degrading the NIRSpec data to MUSE-like conditions indicates that differences in PSFs are likely responsible for these differences.

(vi) The ALMA CO(2--1) observations reveal no extended molecular gas counterpart to the filamentary structures seen in \paa and \hs{1}, with detections limited to faint and compact clumps. This suggests a spatial segregation between cold molecular gas and warmer gas phases in the nuclear region.

(vii) The stellar kinematics show a weak velocity gradient below the instrumental resolution, whose orientation from southeast to northwest is coherent between NIRSpec and MUSE observations, and exhibit a smooth, centrally peaked velocity dispersion profile. In contrast, the gas displays complex and spatially structured kinematics, indicating that the stellar and gaseous components are dynamically decoupled. Similarly, the ICM probed by XRISM presents vastly different bulk velocities than these obtained with our JWST data, further suggesting dynamic decoupling.

(viii) We derive systemic redshift estimates from both stellar and gaseous tracers. Using the mean stellar velocity, we obtain $z = 0.00984(5)$, while the mean heliocentric velocity of the \paa emission in the \textit{Filament C} region yields $z = 0.00999(2)$. The slight offset between these values is consistent with the complex kinematics of the gas relative to the stellar component.

Taken together, these results support a scenario in which the gas in NGC~4696 is dynamically complex and multi-phase, with inflowing filamentary structures feeding the central regions. These observations highlight the unique capability of JWST/NIRSpec to resolve gas kinematics on sub-kiloparsec scales in BCGs, enabling direct investigation of accretion processes that are inaccessible to ground-based facilities. Future JWST observations of NGC~4696 will be essential to further constrain the kinematics of gas in similar BCGs, like NGC 1275.

\begin{acknowledgments}

The authors wish to thank the anonymous referee for their helpful comments as well as \'Etienne Artigau for sharing with us his knowledge of stellar populations. MM acknowledges the support of the Natural Sciences and Engineering Research Council of Canada (NSERC) and of the Fonds de recherche du Qu\'ebec (FRQ) (\href{https://doi.org/10.69777/365532}{doi: 10.69777/365532}). JHL acknowledges funding support from the Canada Research Chairs Program, as well as NSERC through the Discovery Grant, Accelerator Supplement programs and the Arthur B. McDonald Fellowship. In addition, this research was made possible through funding from the Canadian Space Agency (GRANT 24JWGO3A06). OP acknowledges the support of NSERC. MR acknowledges support from the National Science Foundation Graduate Research Fellowship under Grant No. 2141064. J. B-B acknowledges support from project UNAM DGAPA-PAPIIT AG 101025, Mexico, as well as support from the PASPA 2025 grant. LA acknowledges support from the Canadian Space Agency through grant 22EXPJWST. MP acknowledges funding from the Physics Department of the University of Montreal (UdeM) and the Centre for research in astrophysics of Qu\'ebec (AstroQu\'ebec). MLGM acknowledges financial support from NSERC via the Discovery grant program and the Canada Research Chair program. YL acknowledges support from NASA grant 80NSSC22K0668, Chandra X-ray Observatory grant TM3-24005X, NSF grant AST-2510198, and CAREER award AST-2516092. VO acknowledges support from the DICYT ESO-Chile Comite Mixto PS 1757, Fondecyt Regular 1251702, and CIRAS-AI Project, code FIUF137139-USACH. This work was supported by grant \url{https://doi.org/10.69777/377645} from the Fonds de recherche du Qu\'ebec and by AstroQu\'ebec.

This work is based in part on observations made with the NASA/ESA/CSA James Webb Space Telescope. The data were obtained from the Mikulski Archive for Space Telescopes at the Space Telescope Science Institute, which is operated by the Association of Universities for Research in Astronomy, Inc., under NASA contract NAS 5-03127 for JWST. These observations are associated with program \#5354. The specific observations analyzed can be accessed via \dataset[doi: 10.17909/yqjb-gg89]{https://doi.org/10.17909/yqjb-gg89}.

This paper makes use of the following ALMA data: ADS/JAO.ALMA\#2016.1.01117.S. ALMA is a partnership of ESO (representing its member states), NSF (USA) and NINS (Japan), together with NRC (Canada), NSTC and ASIAA (Taiwan), and KASI (Republic of Korea), in cooperation with the Republic of Chile. The Joint ALMA Observatory is operated by ESO, AUI/NRAO and NAOJ.

The Universit\'e de Montr\'eal recognizes that it is located on unceded (no treaty) Indigenous territory, and wishes to salute those who, since time immemorial, have been its traditional custodians. The University expresses its respect for the contribution of Indigenous peoples to the culture of societies here and around the world. The Universit\'e de Montr\'eal is located where, long before French settlement, various Indigenous peoples interacted with one another. We wish to pay tribute to these Indigenous peoples, to their descendants, and to the spirit of fraternity that presided over the signing in 1701 of the Great Peace of Montr\'eal, a peace treaty founding lasting peaceful relations between France, its Indigenous allies and the Haudenosauni Confederacy (pronounced: O-di-no-sho-ni). The spirit of fraternity that inspired this treaty is a model for our academic community.

\end{acknowledgments}
\facilities{JWST (NIRSpec), MUSE (WFM), HST (WFC3).}

\software{
    GraphingLib \citep{gustave_coulombe_2026_19936793},
    NumPy \citep{harris_array_2020},
    Astropy \citep{theastropycollaboration_astropy_2013,theastropycollaboration_astropy_2018,theastropycollaboration_astropy_2022},
    pvextractor \citep{ginsburg_radio_2015},
    reproject \citep{robitaille_reproject_2020},
    Matplotlib \citep{hunter_matplotlib_2007},
    palettable \citep{davis_palettable_2022}.
}

\bibliography{sample701}{}
\bibliographystyle{aasjournalv7}

\appendix
\vspace{-0.7cm}
\section{\texorpdfstring{\ho{2}, O(3), Q(7) and Q(9) Emission Line Maps}{O(2), O(3), Q(7) and Q(9) Emission Line Maps}}\label{app:gas_lines}
\vspace{-0.5cm}

\begin{figure}[h!]
    \centering
    \includegraphics[width=0.719\linewidth]{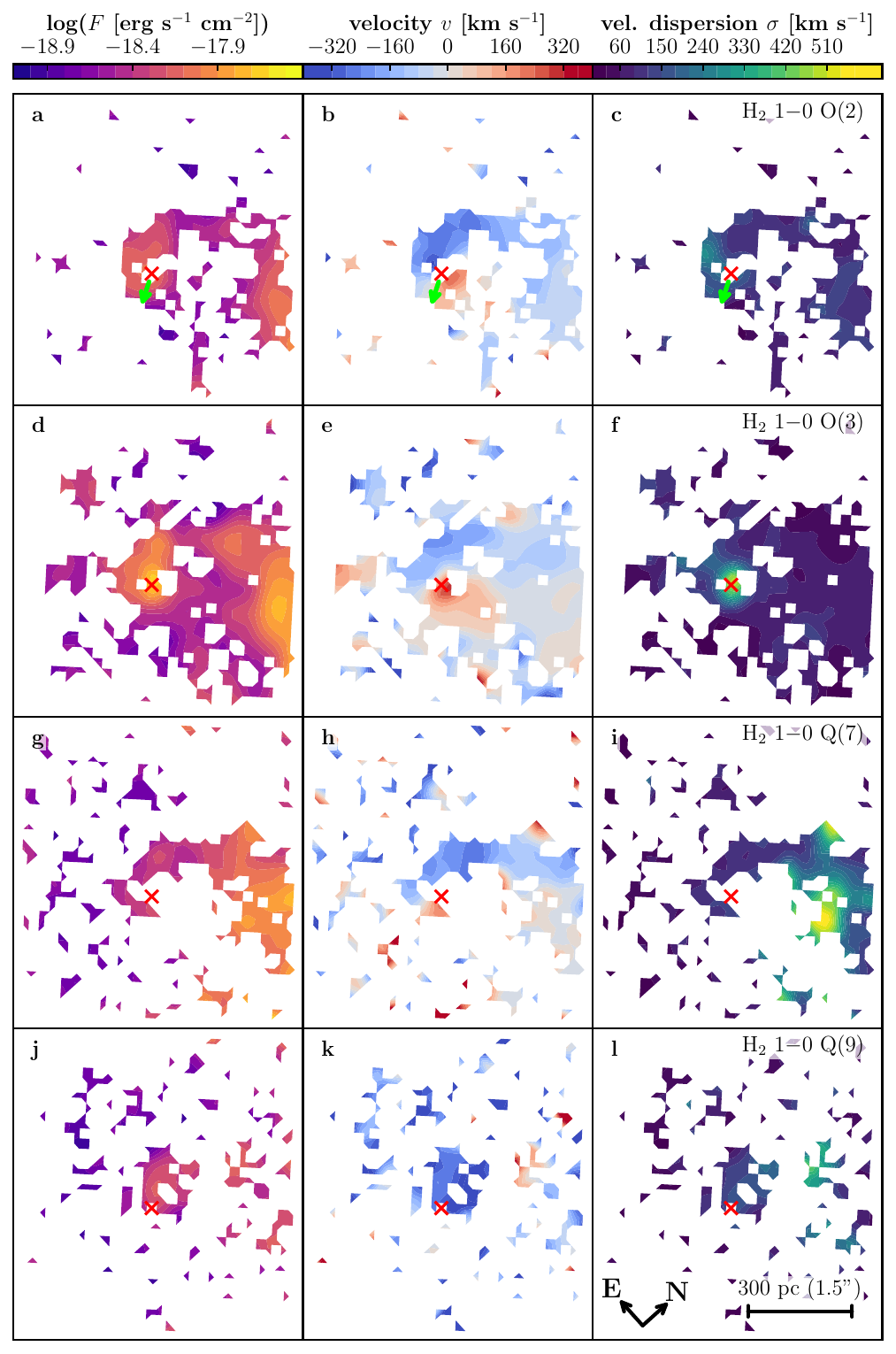}
    \caption{Left to right: logarithmic flux, velocity, and velocity dispersion maps of the \ho{2}, O(3), Q(7) and Q(9) emission lines obtained by fitting the JWST/NIRSpec data using LOKI with a single component. These maps were processed using the methods detailed in Section~\ref{sec:Gas Kinematics}. The red cross marks the position of the kinematic centre (see Section~\ref{sec:Black Hole Position}) and the green arrow in the top row represents the direction of the jet \citep{taylor_lowpower_2006}. All shown emission has SNR $>2$.}
    \label{fig:additional_gas_lines}
\end{figure}

\section{\texorpdfstring{\paa Channel Maps}{Pa alpha Channel Maps}}\label{app:Pa alpha Channel Maps}

\begin{figure}[h!]
    \centering
    \includegraphics[width=\linewidth]{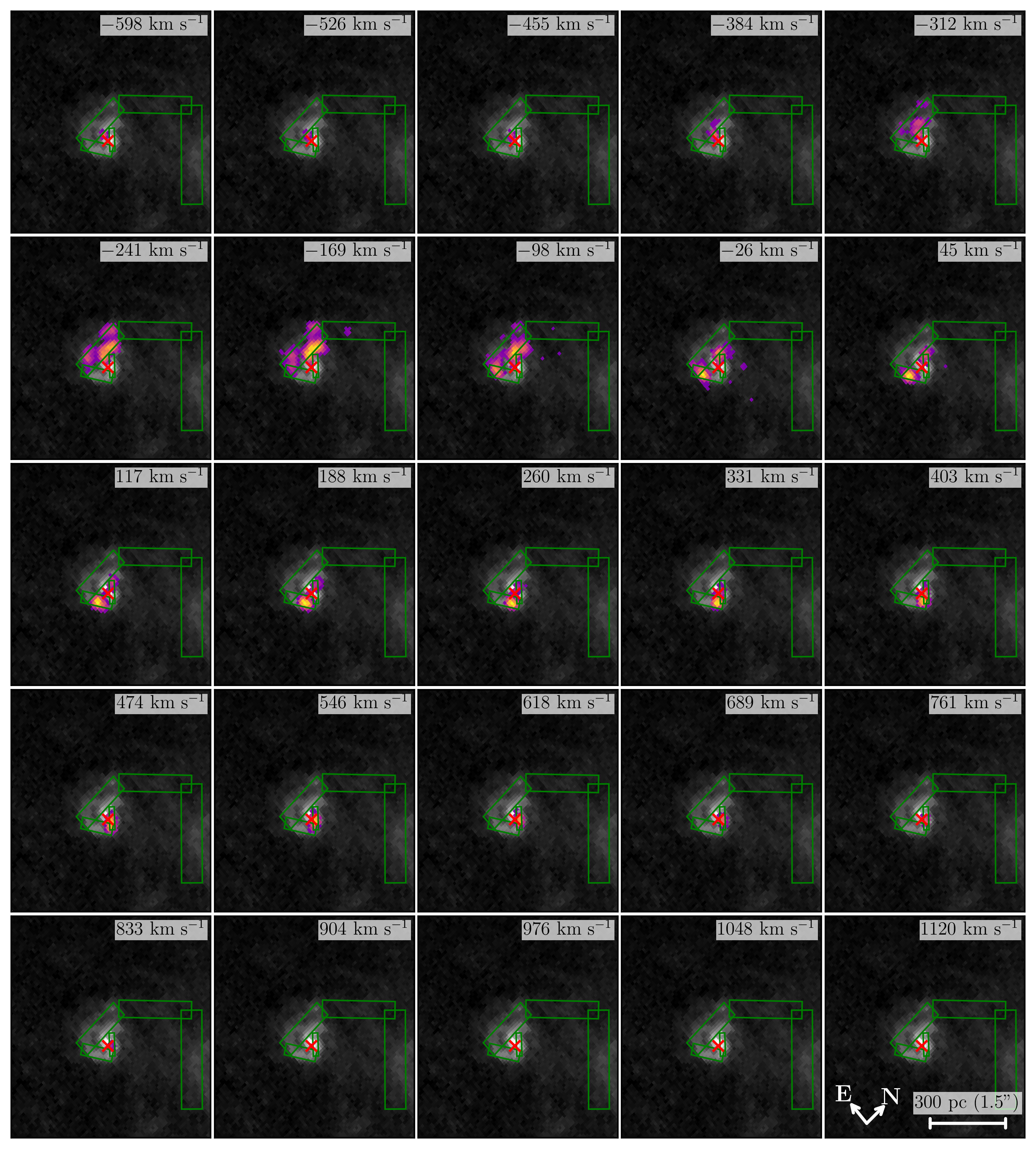}
    \caption{\paa flux at different velocities, showing the velocity gradient along the \textit{Swirl}. This is the channel map version of the animated figure available in the HTML version of this paper. These maps were obtained following the process described in Fig.~\ref{fig:Pa_alpha_gif}. The background image is the HST F665N \ha image subtracted by the F814W filter and the green apertures represent the suggested gas channels (same as in Fig~\ref{fig:PV_inflow}). The red cross marks the position of the kinematic centre (see Section~\ref{sec:Black Hole Position}).}
    \label{fig:pa_alpha_gif_channel_map}
\end{figure}

\section{\texorpdfstring{\hs{1} Triple-Component Emission Profile}{H2S1 Triple-Component Emission Profile}}\label{app:H2S1 Triple-Component Emission Profile}

\begin{figure}[h!]
    \centering
    \includegraphics[width=\linewidth]{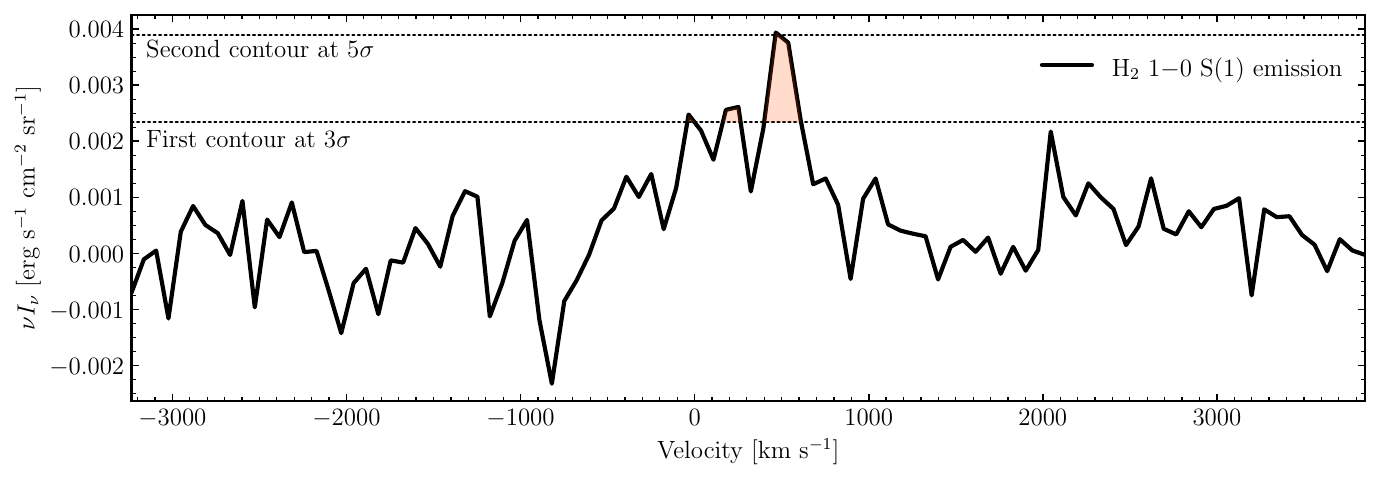}
    \caption{\hs{1} spectrum extracted along the filament after the kinematic centre as shown in Fig.~\ref{fig:PV_inflow}. This profile has been continuum-subtracted and shows three $>3\sigma$ peaks which correspond to the three clumps seen in the last aperture of Fig.~\ref{fig:PV_inflow} \textbf{c} after the kinematic centre (red line). A broad wavelength range is chosen in order to show that these peaks are genuinely statistically significant. The horizontal lines are drawn to match the first two contour levels in Fig.~\ref{fig:PV_inflow}.}
    \label{fig:appendix_h2s1_emission}
\end{figure}



\end{CJK}
\end{document}